\documentclass[a4paper, amsfonts, amssymb, amsmath, reprint, showkeys, nofootinbib, twoside, superscriptaddress, longbibliography]{revtex4-2}
\usepackage[table]{xcolor} 
\usepackage[english]{babel}
\usepackage[utf8]{inputenc}
\usepackage[colorinlistoftodos, color=green!40, prependcaption]{todonotes}
\usepackage[pdftex, pdftitle={Article}, pdfauthor={Author}]{hyperref} 
\usepackage{amsthm}
\usepackage{mathtools}
\usepackage{physics}
\usepackage{graphicx}
\usepackage[left=18mm,right=18mm,top=25mm,columnsep=15pt, bottom=20mm]{geometry}
\usepackage{adjustbox}
\usepackage{placeins}
\usepackage[T1]{fontenc}
\usepackage{lipsum}
\usepackage{csquotes}
\usepackage{xfrac}
\usepackage{enumitem}
\usepackage{array}
\usepackage{multirow}

\begin{document}

\title{Impact of noise transients on gravitational-wave burst detection efficiency of the \textit{BayesWave} pipeline with multi-detector networks}

\author{Yi Shuen C. Lee}
    \email[]{ylee9@student.unimelb.edu.au}
    \affiliation{School of Physics, The University of Melbourne, Victoria 3010, Australia.}
    \affiliation{Australia Research Council Centre of Excellence for Gravitational Wave Discovery (OzGrav).} 
\author{Margaret Millhouse}
    \email[]{meg.millhouse@gatech.edu}
    \affiliation{School of Physics, The University of Melbourne, Victoria 3010, Australia.}
    \affiliation{Australia Research Council Centre of Excellence for Gravitational Wave Discovery (OzGrav).} 
    \affiliation{Center for Relativistic Astrophysics, Georgia Institute of Technology, Atlanta, GA 30332, USA.}
\author{Andrew Melatos}
    \email[]{amelatos@unimelb.edu.au}
    \affiliation{School of Physics, The University of Melbourne, Victoria 3010, Australia.}
    \affiliation{Australia Research Council Centre of Excellence for Gravitational Wave Discovery (OzGrav).}

\begin{abstract}
Detection confidence of the source-agnostic gravitational-wave burst search pipeline \textit{BayesWave} is quantified by the log signal-versus-glitch Bayes factor, $\ln\mathcal{B}_{\mathcal{S},\mathcal{G}}$. A recent study shows that $\ln\mathcal{B}_{\mathcal{S},\mathcal{G}}$ increases with the number of detectors. However, the increasing frequency of non-Gaussian noise transients (glitches) in expanded detector networks is not accounted for in the study. Glitches can mimic or mask burst signals resulting in false alarm detections, consequently reducing detection confidence. This paper an empirical study on the impact of false alarms on the overall performance of \textit{BayesWave}, with expanded detector networks. The noise background of \textit{BayesWave} for the Hanford-Livingston (HL, two-detector) and Hanford-Livingston-Virgo (HLV, three-detector) networks are measured using a set of non-astrophysical background triggers from the first half of Advanced LIGO and Advanced Virgo's Third Observing Run (O3a). Efficiency curves are constructed by combining $\ln\mathcal{B}_{\mathcal{S},\mathcal{G}}$ of simulated binary black hole signals with the background measurements, to characterize \textit{BayesWaves}'s detection efficiency as a function of the per-trigger false alarm probability. The HL and HLV network efficiency curves are shown to be similar. A separate analysis finds that detection significance of O3 gravitational-wave candidates as measured by \textit{BayesWave} are also comparable for the HL and HLV networks. Consistent results from the two independent analyses suggests that the overall burst detection performance of \textit{BayesWave} does not improve with the addition of Virgo at O3a sensitivity, because the increased false alarm probability offsets the advantage of higher $\ln\mathcal{B}_{\mathcal{S},\mathcal{G}}$.

\end{abstract}

\maketitle

\section{Introduction} \label{sec:introduction}

The Advanced Laser Interferometer Gravitational-Wave
Observatory (LIGO) \cite{2015CQGra..32g4001L} detectors in Hanford, Washington and Livingston, Louisiana, USA have completed three observing runs O1, O2 and O3 between 2015 and 2020, two of which were joint observations with the Advanced Virgo detector in Cascina, Italy \cite{2015CQGra..32b4001A}. The Kamioka Gravitational Wave Detector (KAGRA) \cite{KAGRA1, KAGRA2, KAGRA3} located in Hida, Japan also came online towards the end of O3, conducting a joint observation (O3GK) \cite{O3GK} with the GEO600 \cite{GEO600} detector in Hannover, Germany. As of the three observing runs, around 90 candidate gravitational wave (GW) events were collectively observed and reported in the Gravitational-wave Transient Catalogs (GWTCs) \cite*{2019PhRvX...9c1040A, 2021PhRvX..11b1053A, GWTC2.1, 2021arXiv211103606T}. In May 2023, the LIGO-Virgo-KAGRA (LVK) collaboration began the fourth observing run O4 with the two LIGO detectors. The Virgo and KAGRA detectors are also expected to join O4 at a later date.

GW events observed so far by the LVK detectors are compact binary coalescences (CBCs), namely the mergers of binary black holes (BBH), binary neutron stars and neutron star-black hole binaries. CBCs are transient GW events, otherwise known as GW burst sources. Aside from CBCs, we expect to observe GW bursts from other astrophysical sources including but not limited to core-collapse supernovae \cite{2011LRR....14....1F, 2020PhRvD.101h4002A}, pulsar glitches \cite{2014ApJ...787..114S}, magnetar bursts\footnote{Magnetar bursts are short bursts ($\sim 0.1 s$) of soft gamma-rays emitted by highly magnetised, isolated neutron stars. Their physical mechanism is unknown.} \cite{2015SSRv..191..315M, 2019ApJ...874..163A}, nonlinear gravitational memory due to low-mass BBH mergers \cite{2020PhRvD.101j4041E} and cosmic string cusps or kinks \cite{1990PhRvD..42..354S, 2005PhRvD..71f3510D, 2021PhRvL.126x1102A}. In addition, the possibility exists of GW bursts from astrophysical objects or processes that have not yet been discovered through electromagnetic observations. By their nature, GW waveforms of such novel signals are unclassified at present. 

Traditionally, GW transient search pipelines use a matched filter \cite{2012PhRvD..85l2006A, 2012PhRvD..86h4017B, 2014PhRvD..89h4041A, 2017arXiv170501845D} to compare the data to a bank of waveform templates obtained through various waveform modelling techniques \cite{2014PhRvD..89f1502T, 2014LRR....17....2B, 2014PhRvD..89b4003P, 1994PhRvD..49.1707D}. Unlike CBCs, the waveforms of most prospective GW burst sources vary unpredictably from one event to the next and involve complicated physics beyond general relativity (e.g. hydrodynamics and neutrino transport). It is challenging to construct robust models with a few well-defined parameters which predict the waveforms, so template-based matched-filter searches for unmodelled GW bursts are impractical. 

Several developed and emerging pipelines exist to perform source-agnostic GW burst searches \cite{oLIB, XPipeline, MLy}, including but not limited to coherent WaveBurst (cWB) and \textit{BayesWave} (BW). The cWB \cite{2008CQGra..25k4029K, 2016PhRvD..93d2004K, 2021SoftX..1400678D, cWB_wavescan} burst search pipeline is used for offline analysis and online, low-latency generation of triggers for electromagnetic followups. Detection statistics of the cWB algorithm scales with the excess power in the time-frequency domain. BW uses the transdimensional Reversible Jump Markov Chain Monte Carlo (RJMCMC) algorithm which adjusts the model dimension in response to the data. For that reason, BW is computationally intensive and is only used to follow-up potential GW candidates identified by other search pipelines. In the all-sky GW burst searches of the three Advanced LIGO and Advanced Virgo observing runs \cite{2017PhRvD..95d2003A, 2019PhRvD.100b4017A, 2021PhRvD.104l2004A}, cWB is used to analyse the full dataset and BW is used to follow-up cWB triggers \cite{2009PhRvD..80f3007L, 2015PhRvD..91h4034L, 2015CQGra..32m5012C, 2021PhRvD.103d4006C, BW_Repo}. Previous studies have shown that hierarchical implementation of cWB and BW enhances detection confidence \cite{2016PhRvD..93b2002K}. 

 As of O4, the LVK global network comprises of four large-scale detectors. With the commissioning of LIGO-India well under way \cite{LIGOIndia}, the network of GW detectors is expected to expand in the coming years. The expanding network of detectors with improved sensitivities increases the duty cycle, sky coverage and the accuracy of sky localisation \cite{LVK_network, 2011CQGra..28l5023S}. However, having more detectors also increases the susceptibility of the network to transient non-astrophysical disturbances, as noted in O3 \cite{2021PhRvX..11b1053A, 2021arXiv211103606T, 2021PhRvD.104l2004A, 2022CQGra..39x5013D}. These non-Gaussian instrumental noise transients, otherwise known as ``glitches'', appear as excess power in detector data and can mimic or mask unmodelled GW bursts. To enable high confidence detections with high astrophysical significance, glitches have to be identified and mitigated appropriately. Several efforts have been made to identify and characterise glitches by their origin and/or morphology \cite{2017CQGra..34f4003Z, 2018CQGra..35o5017P, 2020SoftX..1200620R, 2020PhRvD.101j2003C, 2023CQGra..40f5004G}. Three common glitches in the LIGO-Virgo detectors are termed blip \cite{2019CQGra..36o5010C}, whistle \cite{2015CQGra..32x5005N} and scattered light \cite{2021CQGra..38b5016S}. The whistle and scattered light glitches are of relatively longer duration ($\sim0.7$-$2.0$ s) and their origins are well-understood. Blip glitches, on the other hand, are transient power spikes which lasts for $\sim 0.1$ s and spans a wide frequency band ($\sim10^2$ Hz), typically of unknown origin. In cases where the glitch origin is unknown, further investigations are necessary before flagging a glitch and regressing it from the data to avoid overlooking astrophysical signals \cite{2012CQGra..29u5008D, 2015CQGra..32p5014T, 2014CQGra..31j5014M, 2019PhRvD..99d2001D, 2019CQGra..36e5011D, 2020PhRvD.101d2003V, 2020PhRvR...2c3066O, galaxies10010012, 2022CQGra..39x5013D}. 

The BW algorithm enables the joint detection and characterisation of GW burst and instrumental glitches, with no \textit{a priori} assumptions of the source or morphology. Studies have been conducted to evaluate various aspects of BW's performance with multi-detector networks, including detection confidence, parameter estimation and waveform reconstruction \cite{2016PhRvD..93b2002K, 2016PhRvD..94d4050L, 2021PhRvD.103f2002L, 2017ApJ...839...15B, 2020PhRvD.102f4056G}. In Ref. \cite{2021PhRvD.103f2002L}, the detection confidence of BW with multi-detector networks is quantified using the algorithm's detection statistic: the log signal-to-glitch Bayes factor, $\ln\mathcal{B}_{\mathcal{S},\mathcal{G}}$. The study showed analytically that increasing the number of detectors in a network has a positive impact on $\ln\mathcal{B}_{\mathcal{S},\mathcal{G}}$, following derivations in Ref. \cite{2016PhRvD..94d4050L}. The results are verified empirically with simulated BBH signals. While the outcome is promising, the study does not consider the increase in glitch rate in an expanded detector network, i.e. it only focuses on the $\ln\mathcal{B}_{\mathcal{S},\mathcal{G}}$ of astrophysical events injected into simulated data in the absence of glitches. This paper generalises Ref. \cite{2021PhRvD.103f2002L}, presenting a fuller analysis of BW's burst detection performance with expanded detector networks by accounting for the detector noise background using real detector data. For noise background measurements, we combine data from the first half of O3 (O3a) for the LIGO Hanford (H), LIGO Livingston (L) and Virgo (V) detectors, in particular the HL (two-detector) and HLV (three-detector) networks. We compare the overall performance of BW between the HL and HLV networks in O3a, noting that Virgo is less sensitive than HL; in contrast, the sensitivities of all three detectors may be comparable in future observing runs. The performance of BW is evaluated by comparing the $\ln\mathcal{B}_{\mathcal{S},\mathcal{G}}$ produced by astrophysical signals against the respective detector network backgrounds, using two independent injection sets. A set of simulated BBH signals is used to construct efficiency curves for characterising BW's detection efficiency as a function of detection significance. To check for consistency, we analyse O3-like CBC signals to measure BW's detection significance of O3 GW candidates from GWTC-2 \cite{2021PhRvX..11b1053A} and GWTC-3 \cite{2021arXiv211103606T}. 

The rest of this paper is organised as follows. In Section \ref{sec:BW} we outline the key features of the BW algorithm. In Section \ref{sec:methods} we discuss the datasets used to study BW's performance: (i) HL and HLV background triggers for background measurements, (ii) simulated BBH injections and (iii) O3-like CBC waveform injections. In Section \ref{sec:backgroundnoise_BW} we present BW's background measurements. In Section \ref{sec:results_IS1} we present the results for BW's efficiency analysis with the simulated BBH injections, and in Section \ref{sec:results_IS2} the significance measurements for the O3 GW candidates. We summarize our findings and discuss avenues for future work in Section \ref{sec:conclusion}.

\section{BayesWave} \label{sec:BW}

In this section, we briefly overview the fundamental principles of the BW algorithm (Section \ref{sec:BWalgorithm_overview}), the models for data reconstruction (Section \ref{subsec:BWmodel}) and the Bayes factor for model selection (Section \ref{sec:BayesianModelSelection}). 

\subsection {Algorithm overview}
\label{sec:BWalgorithm_overview}

The BW algorithm is designed to adaptively reconstruct of non-stationary and non-Gaussian transients in the data, using models with variable dimensions. The name of the algorithm, \textit{BayesWave}, expresses two key concepts: (i) waveform reconstruction using sine-Gaussian (also known as Morlet-Gabor) wavelets, and (ii) the implementation of Bayesian inference to discriminate signals from glitches.

For a given detector $i$, the data $d_i(t)$ at time $t$ consists of three components: the GW signal $h_i(t)$, which is bounded in $t$ for burst sources; glitches $g_i(t)$, which are also bounded in $t$; and random detector noise $n_i(t)$, which is present continuously. That is, we have $d_i(t) = h_i(t) + g_i(t) + n_i(t)$. The BW algorithm attempts to reconstruct the transient, non-Gaussian features i.e. $h_i(t)$ and/or $g_i(t)$ in a stretch of detector data, by summing a set of sine-Gaussian wavelets. A single sine-Gaussian wavelet in the time domain takes the mathematical form
\begin{equation}
    \label{eq:wavelet}
    \Psi(t;\vb*{\lambda}) = Ae^{-(t-t_0)^2 / \tau ^2}\cos\left[2\pi f_0(t-t_0)+\phi_0\right],
\end{equation}
with $\tau=Q/(2\pi f_0)$ and $\vb*{\lambda}=\{t_0, f_0, Q, A, \phi_0\}$. The symbols $t_0, f_0, Q, A, \phi_0$ denote the central time, central frequency, quality factor, amplitude and phase offset of the wavelet respectively. Since the wavelets are not linearly independent, they form a frame and not a basis (see Section 3 of Ref. \cite{2015CQGra..32m5012C} for further details). 

Wavelet parameters are sampled from designated prior distributions using the trans-dimensional Reversible Jump Markov Chain Monte Carlo (RJMCMC) technique \cite{RJMCMC}. The implementation of trans-dimensional jumps allows for the number of wavelets to vary depending on the waveform complexity. By summing all the wavelets at each iteration of the RJMCMC chain, we obtain a posterior distribution of waveform models. For further details on wavelet parameter estimation and the measures taken to optimise convergence to the target distribution, we refer the reader to Refs. \cite{2015CQGra..32m5012C} and \cite{2021PhRvD.103d4006C}.

\subsection{Modelling the data}
\label{subsec:BWmodel}

 BW reconstructs the detector data using three independent models, namely (i) the GW signal plus Gaussian-noise model, $\mathcal{S}$, (ii) glitches plus Gaussian-noise model, $\mathcal{G}$ (iii) Gaussian-noise model, $\mathcal{N}$. In this work, we are interested in the Bayes factor between the $\mathcal{S}$ and $\mathcal{G}$ models as a quantitative measure for BW's detection confidence. 
 
\subsubsection{Signal model, $\mathcal{S}$}
\label{sec:signal_model}
Recall that the five intrinsic parameters of a sine-Gaussian wavelet can represented with a single parameter vector $\vb*{\lambda} = \{t_0, f_0, Q, A, \phi_0\}$. If a real GW signal is present in the data of a multi-detector network, we expect it to be coherent across all detectors in the network, albeit with different signal-to-noise ratio (SNR) and polarization per detector depending on the sensitivity and orientation of the detectors respectively. Therefore when reconstructing the data using the signal model, the same wavelet parameters are used across all detectors in the network. The set of intrinsic parameters for the signal model ($\mathcal{S}$) is given by $\vb*{\lambda}^{\mathcal{S}} = \vb*{\lambda}_1 \cup \vb*{\lambda}_2 \dots \cup \vb*{\lambda}_{N^{\mathcal{S}}}$, where $N^{\mathcal{S}}$ denotes the number of wavelets used in the signal reconstruction. These parameters are geocentric, meaning they are measured at a reference point located at the center of the Earth.
 
 Since the signal models represent astrophysical GW signals, all $N^{\mathcal{S}}$ wavelets used in the reconstruction also share a set of extrinsic parameters $\vb*{\Omega} = \{\theta, \phi, \epsilon, \psi\}$. The symbols denote the right ascension, declination, ellipticity and polarization angle of the GW in order of appearance. The complete set of signal model parameters is then given by $\vb*{\theta}^\mathcal{S} = \vb*{\lambda}^{\mathcal{S}} \cup \vb*{\Omega}$.
 
The geocentrically measured signal waveforms, parameterised by $\vb*{\lambda}^{\mathcal{S}}$, can be projected onto the $i$-th detector using the the detector's unique time delay operator $\Delta t_i(\theta,\phi)$, along with the antenna beam pattern response functions $F^+_i(\theta, \phi, \psi)$ and $F^+_i(\theta, \phi, \psi)$ of the plus ($+$) and cross ($\times$) polarizations\footnote{Antenna pattern functions are typically a function of time. However, the time dependence is omitted here with the assumption that the antenna patterns are constant over the short duration of GW burst. This assumption is conventional across all burst searches.}. Mathematically we write \cite{2011LIGOT010110-00-Z, 2016PhRvD..94d4050L}
 \begin{equation}
 \label{eq:detector_waveform}
     h_i(f; \vb*{\lambda}^\mathcal{S}, \vb*{\Omega}, N^\mathcal{S}) = \left(F^+_i\Tilde{h}_+ +  F^\times_i\Tilde{h}_\times\right) e^{2\pi if \Delta t_i},
 \end{equation}
where $\Tilde{h}_p$ denotes the Fourier transform of the time domain geocentric GW signal, $h_p(t)$ for polarization $p$. The version of BW used in our analysis assumes elliptical polarization such that the ellipticity parameter $\epsilon$ maps $\Tilde{h}_+$ to the cross polarization $\Tilde{h}_\times$ via 
 \begin{equation}
 \Tilde{h}_\times = \epsilon \Tilde{h}_+e^{i\pi/2}, 
 \end{equation}
and $\Tilde{h}_+$ is expressed as a linear combination of sine-Gaussian wavelets in the frequency domain (obtained by taking the Fourier transform of Equation \ref{eq:wavelet}):
 \begin{equation}
     \Tilde{h}_+(f) = \sum^{N^{\mathcal{S}}}_{n=1} \Tilde{\Psi}(f;\vb*{\lambda}_n).
 \end{equation}

\subsubsection{Glitch model, $\mathcal{G}$}
\label{sec:glitch_model}
Unlike GW signals, instrumental glitches and noise are uncorrelated across the detector network. Therefore the glitch model uses independent sets of wavelets to reconstruct glitches in each detector. Let $N^{\mathcal{G}_i}$ denote the number of wavelets and $\vb*{\lambda}^{\mathcal{G}_i} = \vb*{\lambda}_{1}^i \cup \vb*{\lambda}_{2}^i \dots \cup \vb*{\lambda}_{N^{\mathcal{G}_i}}^i$ be the set of wavelet parameters used in the glitch model reconstruction of detector $i$. We can then write the glitch model for the $i$-th detector as
\begin{equation}
    g(\vb*{\lambda}^{\mathcal{G}_i} , N^{\mathcal{G}_i}) = \sum^{N^{\mathcal{G}_i}}_{n=1} \Tilde{\Psi}(f; \vb*{\lambda}^{i}_n).
\end{equation}
Thus for a network with $\mathcal{I}$ detectors, the complete set of glitch model wavelet parameters is given by $\vb*{\theta}^{\mathcal{G}}= \vb*{\lambda}^{\mathcal{G}_1} \cup \vb*{\lambda}^{\mathcal{G}_2} \dots \cup \vb*{\lambda}^{\mathcal{G}_\mathcal{I}}$. Note that there are no extrinsic parameters in the glitch model as it assumes the non-Gaussianity in the data to be independent in each detector (i.e. non-astrophysical) \cite{2016PhRvD..94d4050L}. 

\subsubsection{Gaussian-noise model, $\mathcal{N}$}
In contrast to $\mathcal{S}$ and $\mathcal{G}$ which models non-Gaussian transient components of the detector data, the \textit{BayesLine} algorithm \cite{2015PhRvD..91h4034L} is implemented within BW to model the Gaussian-noise power spectral density (PSD). LIGO and Virgo Gaussian noise sources can be classified into three broad frequency bands: (i) seismic noise ($\sim 10$ Hz), (ii) thermal noise ($\sim 10$-$200$ Hz) and (iii) quantum (photon) shot noise ($\gtrsim 200$ Hz). Moreover, various aspects of the detector apparatus including mirror suspensions, calibration lines, and the AC electrical supply are recurrent sources of high-power, narrow-band spectral lines. \textit{BayesLine} collectively characterises these noise features by modelling the PSD using cubic splines and Lorentzians as bases to fit smooth broad-band noise and narrow-band line-like features respectively. The mathematical details of \textit{BayesLine} are incidental to this paper, a full description can be found in \cite{2015PhRvD..91h4034L}.

By amalgamating all plausible PSD models from \textit{BayesLine}, we obtain $\mathcal{N}$. As its name suggests, $\mathcal{N}$ models the data as purely Gaussian noise. In fact $\mathcal{S}$ and $\mathcal{G}$ also incorporate \textit{BayesLine} for PSD estimation on top of the wavelet models for non-Gaussian feature(s), and are therefore known as composite models \cite{2015CQGra..32m5012C}.

\subsection{Bayesian model selection}
\label{sec:BayesianModelSelection}
BW compares the model evidences via the Bayes factor to give the relative odds between the hypotheses described in Section \ref{subsec:BWmodel}. For a given model $\mathcal{M}$, the evidence is calculated by
\begin{equation}
    \label{eq:model_evidence}
    p(\vb*{d}|\mathcal{M}) = \int d\vb*{\theta}^\mathcal{M} p(\vb*{\theta}^\mathcal{M}|\mathcal{M})p(\vb*{d}|\vb*{\theta}^\mathcal{M}, \mathcal{M})
\end{equation}
where $p(\vb*{\theta}^\mathcal{M}|\mathcal{M})$ is the prior i.e the probability that $\mathcal{M}$ is parameterised by $\vb*{\theta}^\mathcal{M}$ prior to observation of the data $\vb*{d}$; and $p(\vb*{d}|\vb*{\theta}^\mathcal{M}, \mathcal{M})$ is the likelihood of observing $\vb*{d}$ given $\vb*{\theta}^\mathcal{M}$. In essence, the evidence is the likelihood of producing the data $\vb*{d}$ from the hypothesis $\mathcal{M}$ marginalised over the parameter space of $\vb*{\theta}^\mathcal{M}$, thus it is otherwise known as the marginalised likelihood \cite{2016PhRvD..94d4050L}. Obtaining model evidences directly from the integral in Equation \ref{eq:model_evidence} is computationally expensive, especially for complex and highly parameterised models. Therefore BW combines the parallel-tempered RJMCMC algorithm \cite{2005PThPS.157..317W} and thermodynamic integration \cite{goggans2004bayesian} to compute the evidences. Implementations of these methods are detailed in Refs. \cite{2015CQGra..32m5012C} and \cite{2015PhRvD..91h4034L}.

The Bayes factor between two models, $\mathcal{M}_\alpha$ and $\mathcal{M}_\beta$, is the ratio of their evidences:
\begin{equation}
    \label{eq:bayesF}
    \begin{split}
    \mathcal{B}_{\alpha, \beta}(\vb*{d}) &= \frac{p(\vb*{d}|\mathcal{M}_\alpha)}{p(\vb*{d}|\mathcal{M}_\beta)}.
    \end{split}
 \end{equation} 
$\mathcal{B}_{\alpha, \beta}(\vb*{d})>1$ suggests that  $\mathcal{M}_\alpha$ is more strongly supported by the data and vice versa. The Bayes factor inherently considers model complexity in model selection by penalising over-fitting. This is a corollary of Occam's razor, which prefers simplicity over complexity amongst competing models. Occam's razor is not deliberately implemented; rather it is an inherent consequence of using Bayes' Theorem and enters via the parameter space volume in Equation \ref{eq:model_evidence}. For a detailed mathematical interpretation, we refer the reader to Section IV.A of \cite{2021PhRvD.103f2002L}.

A study conducted by Littenberg et al. \cite{2016PhRvD..94d4050L} to assess BW's ability to distinguish between GW signals and instrumental glitches shows that for a two-detector network with interferometers of equal sensitivity (i.e. the HL network), the primary scaling of the Bayes factor goes as \cite{2016PhRvD..94d4050L}
\begin{equation}
\label{eq:BSG_scaling}
\ln\mathcal{B}_{\mathcal{S},\mathcal{G}}\propto N \ln(\text{SNR}_\text{net}).
\end{equation}
A simplifying assumption is that the number of wavelets used in the signal model $\mathcal{S}$ is the same as the glitch model $\mathcal{G}$ for a single detector, viz. $N^{\mathcal{S}}=N^{\mathcal{G}_i}=N$. For a network with $\mathcal{I}$ detectors, the overall network SNR of the non-Gaussian transient in the data is given by 
\begin{equation}
\text{SNR}_\text{net}^2 = \sum ^\mathcal{I} _{i=1} \text{SNR}^2_i
\label{eq:SNR_net}
\end{equation}
where SNR$_i$ is the SNR in detector $i$. Altogether, Equation \ref{eq:BSG_scaling} suggests that $\ln\mathcal{B}_{\mathcal{S},\mathcal{G}}$ and hence detection confidence scale with both signal strength and waveform complexity. 
    
In a complementary study \cite{2021PhRvD.103f2002L}, BW is used to recover injected BBH signals from the HL, HLV and HLKV networks to quantify its detection confidence with expanded detector networks. In this study, $\ln\mathcal{B}_{\mathcal{S},\mathcal{G}}$ is further shown to scale with the number of detectors in the network, $\mathcal{I}$, according to
\begin{equation}
\ln\mathcal{B}_{\mathcal{S},\mathcal{G}}\propto \mathcal{I}N \ln(\text{SNR}_\text{net}).
\label{eq:BSG_multi}
\end{equation}
In other words, BW's detection confidence is directly and positively impacted by increasing the number of detectors in the network, all else being equal. 

\section{BayesWave efficiency analysis} \label{sec:methods}

In standard GW searches, the astrophysical significance of a detection candidate is determined by the frequency of false alarms. False alarms are non-astrophysical events with detection statistics corresponding to that of GW candidates. To estimate the prevalence of false alarms, one can count the number of triggers produced by the detector background which does not contain astrophysical signals.
 
Ref. \cite{2021PhRvD.103f2002L} assesses BW's detection confidence for expanded detector networks using only the detection statistic $\ln\mathcal{B}_{\mathcal{S},\mathcal{G}}$ produced by astrophysical events. However as the global detector network expands, the likelihood of instrumental glitches increases. The associated increase in false alarm detections reduces astrophysical significance of detections, thereby reducing detection confidence. Unmodelled burst searches (e.g. with BW) place fewer constraints on the waveform morphology, and are therefore confounded more readily by glitches compared to modelled searches (e.g. with a matched filter) \cite{2008CQGra..25r4004B, 2009PhRvD..80j2001A, LIGO_Phys_report_2009, 2012CQGra..29o5002A, PhysRevD.85.122007}. Since the significance of $\ln\mathcal{B}_{\mathcal{S},\mathcal{G}}$ is influenced by false alarms, we present a more complete analysis of BW's performance with expanded detector networks by considering the impact of detector noise backgrounds on detection confidence.

We use detection efficiency $P_{\rm det}$ as a figure of merit to compare the overall performance of BW between the HL (two-detector) and HLV (three-detector) networks. $P_{\rm det}$ is typically characterised as a function of detection significance by means of a receiver-operating-characteristic (ROC) curve\footnote{Typical ROC curves plot probability of detection (true positives) on the vertical axis and probability of false alarm (false positives) on the horizontal axis}, also known as an efficiency curve. In this study, we use the per-trigger false alarm probability $P_{\rm FA}$ as a measure of significance. We define $P_\text{FA}$ as the probability that a trigger measured with a given detection statistic is a false alarm, and $P_{\rm det}$ as the probability of detecting an astrophysical event with a given significance. Higher $P_{\rm FA}$ indicates low astrophysical significance. Therefore higher $P_{\rm det}$ is achieved, if higher $P_{\rm FA}$ is tolerated. $P_\text{FA}$ should not be confused with the false alarm rate (FAR), which measures the number of false alarms per unit time. We further discuss the difference in Section \ref{sec:backgroundnoise_BW}, and explain why we use $P_\text{FA}$ instead of FAR in our analysis. 

In order to measure $P_\text{FA}$, we need to understand the distribution of $\ln\mathcal{B}_{\mathcal{S},\mathcal{G}}$ produced by the detector noise background. It is challenging to construct models that can accurately predict the noise background, so we empirically obtain the background distribution by applying BW to triggers identified by cWB from time-shifted background data of the HL and HLV networks. The distribution of $\ln\mathcal{B}_{\mathcal{S},\mathcal{G}}$ produced by the background triggers is then used to compute $P_\text{FA}$. Using BW to recover a population of injected signals, we obtain a distribution of $\ln\mathcal{B}_{\mathcal{S},\mathcal{G}}$ for astrophysical events. Combining $\ln\mathcal{B}_{\mathcal{S},\mathcal{G}}$ of background triggers and astrophysical injections, we compute $P_{\rm det}$ as a function of $P_{\rm FA}$ to construct efficiency curves. We discuss the methods of constructing efficiency curves in greater detail in Section \ref{sec:construct_effCurves}. 

To study the impact of the noise background on BW's overall performance with expanded detector networks, we compare the efficiency curves between the HL and HLV networks for a synthesised population of BBHs. As a consistency check, we also analyse a set of O3-like CBC waveforms to measure and compare BW's detection significance of O3 GW detection events for HL and HLV. In the following sections, we detail the background and injection datasets for the analyses.

\subsection{Background data} 
\label{sec:glitch_dataset}

In GW data analysis, it is standard practice to use the time-shifting method to create pseudo-real detector datasets for noise background estimations \cite{2010CQGra..27a5005W, 2010PhRvD..81j2001A, 2016PhRvD..93l2004A, 2017PhRvD..95d2003A, 2019PhRvD.100b4017A, 2021PhRvD.104l2004A}. The time-shifting method introduces artificial time off-sets between the outputs of GW detectors operating in concert. The offsets are much larger than the coherence time ($\sim 10 \, {\rm ms}$) of any real GW signals between the detectors, determined by the distance between the detectors and the GW propagation speed. As a result, coincident triggers in the time-shifted data cannot be astrophysical. By performing time-shifts repetitively on months worth of detector data, we obtain an artificially extended set of background data with effective livetimes\footnote{The extended time interval obtained as a result of time-shifting is known as the effective livetime.} spanning thousands of years. We can use this to estimate $P_{\rm FA}$ by empirically measuring the fraction of background (i.e. noise-induced) triggers above a selected detection threshold in the time-shifted background \cite{2010CQGra..27a5005W}. 

In the Advanced LIGO and Advanced Virgo all-sky searches for short GW bursts \cite{2017PhRvD..95d2003A, 2019PhRvD.100b4017A, 2021PhRvD.104l2004A}, the cWB algorithm is used to analyse the full observational data. Due to the implementation of RJMCMC, BW is computationally intensive. Thus, BW is only used to follow-up subsets of cWB triggers. Although Ref. \cite{2016PhRvD..93b2002K} has shown that the hierarchical implementation of cWB and BW enhances detection confidence in all-sky burst searches, the aim of our study is to assess the independent burst detection performance of BW. By convention, we use pre-existing trigger lists generated by the cWB pipeline \cite{klimenko_sergey_2021_4419902, 2021PhRvD.104l2004A} to downselect triggers for BW background measurements, but we do not make any claims on cWB's background and detection efficiency. We choose to use the trigger list for the first half of O3 (O3a), acquired from the cWB low-frequency ($16$-$1024$ Hz) all-sky analysis of the full time-shifted O3a background data. The analysis is conducted separately for the HL and HLV networks, on background data obtained by applying time-shifts on 104.94-day (HL) and 75.19-day (HLV) segments of the real-time O3a detector data. The time-shifted background data accumulates 981 years and 573 years of effective livetimes for the HL and HLV networks respectively. 

We select triggers by thresholding their cWB detection statistic, $\rho$, which scales with the network SNR of the signal present in the data \cite{2008CQGra..25k4029K, 2016PhRvD..93d2004K}. We arbitrarily nominate $\rho_{\text{threshold}}=7$ as the significance threshold, in line with previous work \cite{2016PhRvD..93b2002K, 2021PhRvD.104l2004A}. Triggers below the threshold are presumed to have insignificant impacts on detection efficiency and therefore excluded from the noise background measurement. cWB identifies $2\times10^3$ and $7\times10^3$ triggers\footnote{These counts include triggers from all three search bins used in the cWB O3a low-frequency burst analysis: LF1, LF2 and LF3. The bins are classified based on trigger morphologies. Classification details can be found in \cite{2021PhRvD.104l2004A} and \cite{2022PhRvD.105f3024L}.} with $\rho>\rho_{\text{threshold}}$ in the HL and HLV background datasets respectively, but thousands of triggers are still too expensive to handle computationally in this paper. A straightforward approach is to increase $\rho_{\text{threshold}}$, but that would deliberately exclude low-SNR triggers from the background measurement. To avoid implementing a stricter $\rho_{\text{threshold}}$, we run BW on a fraction (denoted by $X$) of randomly selected triggers from the full trigger list, all of which satisfy $\rho>\rho_{\text{threshold}}=7$. We set $X=0.45$ and $X=0.15$ for the HL and HLV datasets respectively to deliver roughly equal numbers of triggers from the two networks. The reduced HL and HLV background datasets consist of 1008 and 1134 triggers respectively. We employ BW to analyse the datasets to obtain $\ln\mathcal{B}_{\mathcal{S},\mathcal{G}}$ for each background trigger. The BW analysis uses the same settings as the Advanced LIGO and Advanced Virgo O3 all-sky search for short GW bursts \cite{2021PhRvD.104l2004A} (see Appendix \ref{app:BWsettings}). 

From the BW analysis, we flag background triggers that are more consistent with the pure Gaussian-noise model, $\mathcal{N}$ than the composite signal plus Gaussian-noise model $\mathcal{S}$. By definition, the $\ln\mathcal{B}_{\mathcal{S},\mathcal{N}}$ error bars of these triggers encompass values less than or equal to zero, i.e. 
\begin{equation}
    \ln\mathcal{B}_{\mathcal{S},\mathcal{N}} - \Delta \ln\mathcal{B}_{\mathcal{S},\mathcal{N}} \leq 0 
\end{equation}
where $\Delta \ln\mathcal{B}_{\mathcal{S},\mathcal{N}} = \sqrt{\left[\Delta \ln p(\vb*{d}|\mathcal{S})\right]^2 + \left[\Delta \ln p(\vb*{d}|\mathcal{N})\right]^2}$
is the width of the error bars and $\Delta \ln p(\vb*{d}|\mathcal{M})$ denotes the uncertainty of the log evidence of model $\mathcal{M}$ \cite{2015CQGra..32m5012C}. For these Gaussian-noise-like triggers, the $\ln\mathcal{B}_{\mathcal{S},\mathcal{G}}$ is meaningless as it serves to compare the evidences of models that characterise non-Gaussianity. Nevertheless, these triggers cannot be discarded from the background measurement as they satisfy $\rho>\rho_{\text{threshold}}$. Therefore we assign them with an arbitrarily low detection statistic, $\ln\mathcal{B}_{\mathcal{S},\mathcal{G}}=-500$, to indicate minimal astrophysical significance. A total of 268 (218) out of 1008 (1134) HL (HLV) triggers are assigned $\ln\mathcal{B}_{\mathcal{S},\mathcal{G}}=-500$. 

We present and discuss the background measurements in Section \ref{sec:backgroundnoise_BW}. 

\subsection{Injections}\label{sec:inj}
In addition to the background measurement, the detection statistic distribution for astrophysical signals is required to evaluate BW's burst detection performance. We inject waveforms of known morphology and recover them using BW to empirically measure the distribution of $\ln\mathcal{B}_{\mathcal{S},\mathcal{G}}$. Since CBCs are well-understood, we use them in this study to assess BW's independent performance with HL and HLV. We analyse two different source populations: Injection Set 1 (IS1), comprising phenomenological BBH waveforms with fixed component masses but uniformly distributed SNR and extrinsic parameters, and Injection Set 2 (IS2), comprising CBC waveforms with parameters that resemble real GW events from O3. The following two subsections describe the objective and properties of each injection dataset in order.

\subsubsection{Phenomenological BBH waveforms (IS1)}\label{sec:simulated_inj} 

IS1 consists of simulated BBH waveforms with a choice of parameter space encompassing the range detectable by the Advanced LIGO and Advanced Virgo detectors. The waveforms are added to temporally spread out segments of HL and HLV data across all of O3a to reflect practical observation intervals. We use IS1 to characterise BW's detection efficiency ($P_{\rm det}$) as a function of detection significance ($P_{\rm FA}$) via efficiency curves and compare the performance of BW with the HL and HLV networks. 

IS1 copies the injection set described in Section V of Ref. \cite{2021PhRvD.103f2002L}. It consists of 1200 simulated BBH waveforms phenomenologically modelled using the \texttt{IMRPhenomD} \cite{PhysRevD.93.044006, PhysRevD.93.044007} approximant. The BBH sources are non-spinning, non-precessing and have equal component masses of $30M_\odot$. They also have uniformly distributed sky locations, inclinations and polarisation angles. The distances are randomly sampled such that the signal amplitude is detectable in simulated HLV data with network signal-to-noise ratio within range $10 \leq \rm SNR_{\rm net} \leq 50$. We use the same injection dataset for both the HL and HLV networks; we simply exclude Virgo data in the HL analysis. By Equation \ref{eq:SNR_net}, we expect SNR$_\text{net}$ of any given event to be lower in the HL network compared to HLV.

The analysis in Ref. \cite{2021PhRvD.103f2002L} injects and recovers waveforms using projected (simulated) O4 detector data. However, BW's background measurements for HL and HLV in this study are carried out using O3a background triggers as discussed in Section \ref{sec:glitch_dataset}. In order to measure the $P_{\rm det}$ as a function of $P_{\rm FA}$, detection statistics ($\ln\mathcal{B}_{\mathcal{S},\mathcal{G}}$) of the astrophysical signals must be compared with the background triggers of the same detector data. Thus we inject IS1 into arbitrarily selected segments of HL and HLV data throughout O3a. The O3a strain data is publicly available at the \href{https://www.gw-openscience.org/data/}{Gravitational Wave Open Science Centre (GWOSC)} \cite{2023arXiv230203676T, GWOSC_O3a} and Figure 2 of Ref. \cite{2021PhRvX..11b1053A} shows representative amplitude spectral densities of the detectors. As with the background, IS1 is analysed using the same BW settings as Ref. \cite{2021PhRvD.104l2004A}. 
 
Events of IS1 are injected into O3a data with the same distances sampled from the simulated data. Since O3a data is noisier and has a different characteristic PSD compared to the simulated HLV data, the $\text{SNR}_\text{net}$ of IS1 events when injected into O3a data is lower than the referenced range $10 \leq \rm SNR_{\rm net} \leq 50$. In order to assess BW's performance under conditions relevant to practical searches, events below a designated detection threshold must be eliminated from the injection data. This is because they cannot serve as triggers by definition, in the same way that background triggers with $\rho < \rho_{\rm threshold} = 7$ do not count as false alarms. Since this a designated search to assess the stand-alone efficiency of the BW algorithm, independent of cWB, we set a nominal significance threshold of $\text{SNR}_\text{cut-off}=10$ for BW viz. only injection events with $\text{SNR}_\text{net}\geq\text{SNR}_\text{cut-off}$ in \textit{both} the HL and HLV networks are adequately significant to be followed-up by BW and included in the efficiency curve analysis. Out of 1200 injections, 412 non-detection events are filtered out from IS1, leaving 788 events going forward. 

From the remaining 788 events, BW identifies 157 (89) events consistent with Gaussian noise in the HL (HLV) network according to the $\ln\mathcal{B}_{\mathcal{S},\mathcal{N}}$ constraint defined in Section \ref{sec:glitch_dataset}. These events are retained in the analysis dataset since they satisfy ${\rm SNR}_{\rm net} \geq 10$ but as with the background triggers, they are assigned $\ln\mathcal{B}_{\mathcal{S},\mathcal{G}}=-500$ to indicate low detection significance.

\begin{figure}[t]
\centering
\includegraphics[width=.49\textwidth]{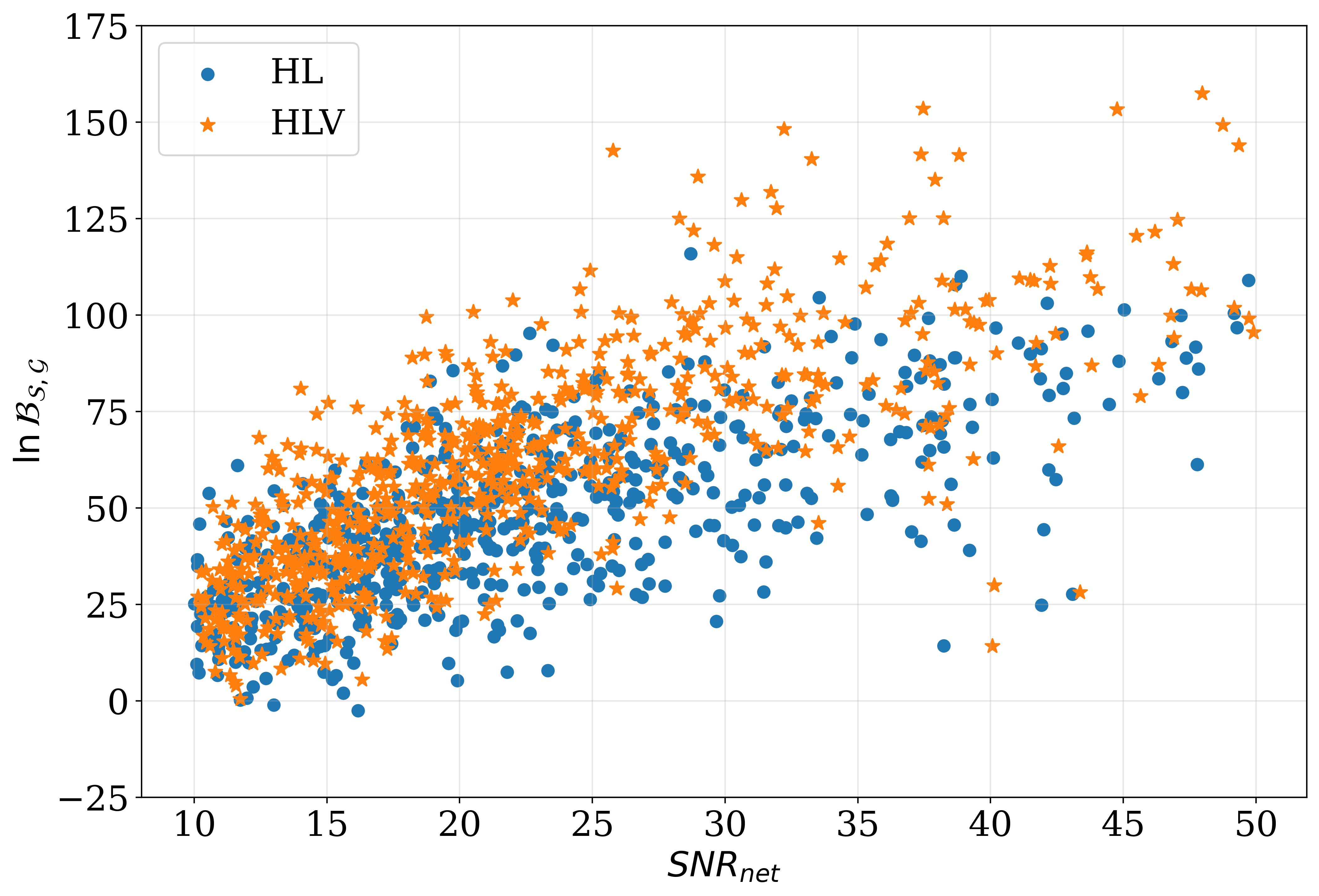}
\caption{Log signal-to-glitch Bayes factor $\ln\mathcal{B}_{\mathcal{S},\mathcal{G}}$ versus network signal-to-noise ratio $\text{SNR}_\text{net}$ for IS1. The blue circles (orange stars) correspond to HL (HLV) network injections; each data point corresponds to a single injection. Gaussian-noise-like events with $\ln\mathcal{B}_{\mathcal{S},\mathcal{G}}=-500$ are not shown.}
    \label{fig:simInj_dist}
\end{figure}

To show the overall distribution of IS1 events, we plot $\ln\mathcal{B}_{\mathcal{S},\mathcal{G}}$ versus $\text{SNR}_\text{net}$ for the HL (blue circles) and HLV (orange stars) injections in Figure \ref{fig:simInj_dist}. The plot shows all but the $\ln\mathcal{B}_{\mathcal{S},\mathcal{G}}=-500$ events to focus on events with astrophysically relevant $\ln\mathcal{B}_{\mathcal{S},\mathcal{G}}$. Injections with comparable $\text{SNR}_\text{net}$ are evidently recovered with higher $\ln\mathcal{B}_{\mathcal{S},\mathcal{G}}$ in HLV compared to HL. This observation is consistent with Ref. \cite{2021PhRvD.103f2002L} where $\ln\mathcal{B}_{\mathcal{S},\mathcal{G}}$ is analytically and empirically shown to increase primarily with $\mathcal{I}$.

Despite the astrophysical origin of IS1, \textit{BayesWave} recovers two of the HL events with $\ln\mathcal{B}_{\mathcal{S},\mathcal{G}}<0\neq -500$ in Figure \ref{fig:simInj_dist}, suggesting that the evidence for the `incoherent' glitch model ($\mathcal{G}$) is higher than for the `coherent' signal model ($\mathcal{S}$). These events are also not consistent with Gaussian-noise i.e. they have $\ln\mathcal{B}_{\mathcal{S},\mathcal{N}}>0$. This is because the injected signal power in the frequency domain is only marginally above the sensitivity threshold in one detector, and is approximately one order of magnitude lower in the other. Equation 2 shows that the sensitivity of each detector to different sky locations, at a given time, depends on the antenna pattern functions. Therefore the $\ln\mathcal{B}_{\mathcal{S},\mathcal{G}}<0$ recovery of the two HL injections, caused by the signal power imbalance across the detectors, is an inadvertent result of mismatched detector sensitivities to the randomly sampled sky locations at the time of injection. With additional coherent signal power from Virgo, the HLV-equivalents of these two events are recovered with $\ln\mathcal{B}_{\mathcal{S},\mathcal{G}}\sim 10^1$. This argument also applies to IS2 injections, discussed in Section \ref{sec:offsource_inj}.

We present the results of BW detection efficiency analysis with IS1 in Section \ref{sec:results_IS1}.

\subsubsection{O3-like CBC waveforms (IS2)}\label{sec:offsource_inj}

To check for consistency with IS1, we measure BW's detection significance for real GW detection events in terms of $P_{\rm FA}$, and compare the measurements between the HL and HLV networks. For this purpose, we implement BW on IS2 consisting CBC waveforms resembling O3a and O3b GW events from GWTC-2 \cite{2021PhRvX..11b1053A} and GWTC-3 \cite{2021arXiv211103606T} respectively, otherwise known as off-source injected waveforms. 

In IS1 the BBH waveforms are sampled from a fixed parameter space and added to detector data spread out across all of O3a; in IS2 the off-source injections are sampled from the matched-filter source parameter posteriors for GW detection events and added into the background data around the event epoch. Off-source injections are used in the GWTCs to test the consistency\footnote{Consistency test are performed by comparing the on-source and off-source match. On-source waveforms are reconstructed directly from the event data. The match, defined by $\mathcal{O} = \langle h_1\mid h_2 \rangle/\sqrt{\langle h_1 \mid h_1 \rangle\langle h_2 \mid h_2 \rangle}$, measures the overlap between two waveforms $h_1$ and $h_2$. $\langle \cdot \mid \cdot \rangle$ is the noise-weighted inner product \cite{PhysRevD.46.5236}. On-source match compares the maximum likelihood waveform from template-based parameter estimation of the actual event with the point estimate from minimally-modelled reconstructions; off-source match compares the off-source injections with their respective reconstructions.} between matched-filter (template-based) CBC waveforms and minimally-modeled waveform reconstructions (e.g. cWB and BW) \cite*{2019PhRvX...9c1040A, 2021PhRvX..11b1053A, 2021arXiv211103606T}. 

IS2 comprises off-source injections of 22 independent GW events detected by the HLV network in O3. We summarise the relevant event properties in Table \ref{table:offsource}. All events listed in Table \ref{table:offsource}, except for GW200202\textunderscore154313, are BW waveform consistency test candidates \cite{2021arXiv211103606T, 2021arXiv211103606T}. GW200202\textunderscore154313 is excluded from the GWTC-3 consistency test due to low on-source match, but since the off-source injections for this event are available there is no reason to exclude it from IS2 for the assessment of BW's detection significance. A set of 200 off-source injections is available for each of the 22 GW events \cite{2021PhRvX..11b1053A, 2021arXiv211103606T}.  We arbitrarily select 50 out of the 200 off-source injections for each GW event, totalling $22\times50=1100$ injections in IS2. Even though a fraction of the GW events are O3b detections, we inject all off-source events into segments of O3a HLV data to ensure comparability with the O3a noise background described in Section \ref{sec:glitch_dataset}. The HL data are equivalent to the HLV data with Virgo removed.

\setlength{\tabcolsep}{0pt}
\rowcolors{2}{gray!15}{white}
\begin{table*}[t]
\begin{tabular}{p{2cm} p{3.5cm}>{\centering}p{2cm}>{\centering}p{2cm}>{\centering}p{2cm}>{\centering\arraybackslash}p{2.5cm}}
\hline
LVK run    & Event name & {$m_1$ \\ ($M_\odot$)}  & {$m_2$\\ ($M_\odot$)}  & Network SNR$^\dag$ & \# off-source injections in IS2 \\ \hline\hline
O3a     & GW190408\_181802 & 24.6$^{+5.1}_{-3.4}$ & 18.4$^{+3.3}_{-3.6}$ & 15.3$^{+0.2}_{-0.3}$ & 50\\ 
O3a     & GW190412         & 30.1$^{+4.7}_{-5.1}$ & 8.3$^{+1.6}_{-0.9}$  & 18.9$^{+0.2}_{-0.3}$ & 48\\ 
O3a     & GW190503\_185404 & 43.3$^{+9.2}_{-8.1}$ & 28.4$^{+7.7}_{-8.0}$ & 12.4$^{+0.2}_{-0.3}$ & 47\\
O3a     & GW190512\_180714 & 23.3$^{+5.3}_{-5.8}$ & 12.6$^{+3.6}_{-2.5}$ & 12.2$^{+0.2}_{-0.4}$ & 40\\ 
O3a     & GW190513\_205428 & 35.7$^{+9.5}_{-9.2}$ & 18.0$^{+7.7}_{-4.1}$ & 12.9$^{+0.3}_{-0.4}$ & 49\\ 
O3a     & GW190517\_055101 & 37.4$^{+11.7}_{-7.6}$ & 25.3$^{+7.0}_{-7.3}$ & 10.7$^{+0.4}_{-0.6}$ & 21\\ 
O3a     & GW190519\_153544 & 66.0$^{+10.7}_{-12.0}$ & 40.5$^{+11.0}_{-11.1}$ & 15.6$^{+0.2}_{-0.3}$ & 48\\ 
O3a     & GW190521         & 95.3$^{+28.7}_{-18.9}$ & 69.0$^{+22.7}_{-23.1}$ & 14.2$^{+0.3}_{-0.3}$ & 44\\ 
O3a     & GW190602\_175927 & 69.1$^{+15.7}_{-13.0}$ & 47.8$^{+14.3}_{-17.4}$ & 12.8$^{+0.2}_{-0.3}$ & 44\\ 
O3a     & GW190706\_222641 & 67.0$^{+14.6}_{-16.2}$ & 38.2$^{+14.6}_{-13.3}$ & 12.6$^{+0.2}_{-0.4}$ & 41\\ 
O3a     & GW190720\_000836 & 13.4$^{+6.7}_{-3.0}$ & 7.8$^{+2.3}_{-2.2}$  & 11.0$^{+0.3}_{-0.7}$ & 24\\ 
O3a     & GW190727\_060333 & 38.0$^{+9.5}_{-6.2}$ & 29.4$^{+7.1}_{-8.4}$ & 11.9$^{+0.3}_{-0.5}$ & 49\\ 
O3a     & GW190728\_064510 & 12.3$^{+7.2}_{-2.2}$ & 8.1$^{+1.7}_{-2.6}$  & 13.0$^{+0.2}_{-0.4}$ & 48\\ 
O3a     & GW190828\_063405 & 32.1$^{+5.8}_{-4.0}$ & 26.2$^{+4.6}_{-4.8}$ & 16.2$^{+0.2}_{-0.3}$ & 48\\ 
O3a     & GW190828\_065509 & 24.1$^{+7.0}_{-7.2}$ & 10.2$^{+3.6}_{-2.1}$ & 10.0$^{+0.3}_{-0.5}$ & 19\\ 
O3a     & GW190915\_235702 & 35.3$^{+9.5}_{-6.4}$ & 24.4$^{+5.6}_{-6.1}$ & 13.6$^{+0.2}_{-0.3}$ & 47\\ 
O3a     & GW190924\_021846 & 8.9$^{+7.0}_{-2.0}$  & 5.0$^{+1.4}_{-1.9}$  & 11.5$^{+0.3}_{-0.4}$ & 36\\ 
\hline
O3b     & GW200129\_065458 & 34.5$^{+9.9}_{-3.2}$ & 28.9$^{+3.4}_{-9.3}$ & 26.8$^{+0.2}_{-0.2}$ & 50\\ 
O3b     & GW200202\_154313 & 10.1$^{+3.5}_{-1.4}$ & 7.3$^{+1.1}_{-1.7}$  & 10.8$^{+0.2}_{-0.4}$ & 35\\ 
O3b     & GW200219\_094415 & 37.5$^{+10.1}_{-6.9}$ & 27.9$^{+7.4}_{-8.4}$ & 10.7$^{+0.3}_{-0.5}$ & 13\\ 
O3b     & GW200224\_222234 & 40.0$^{+6.9}_{-4.5}$ & 32.5$^{+5.0}_{-7.2}$ & 20.0$^{+0.2}_{-0.2}$ & 36\\ 
O3b     & GW200311\_115853 & 34.2$^{+6.4}_{-3.8}$ & 27.7$^{+4.1}_{-5.9}$ & 17.8$^{+0.2}_{-0.2}$ & 48\\ \hline
\end{tabular}
\caption{List of O3 GW events used to generate the off-source injections of IS2. The columns from left to right show: (i) The LIGO-Virgo-KAGRA (LVK) observing run in which the event was detected, (ii) event name, (iii) primary component mass $m_1$, (iv) secondary component mass $m_2$, (v) HLV network matched-filter SNR$^{\dag}$ and (vi) number of off-source injections (out of 50) that satisfy $\text{SNR}_\text{net}\geq\text{SNR}_\text{cut-off}$ and retained in IS2. Source parameter values displayed in the table are the median and the 90\% symmertric credible intervals of the Bayesian posterior. Information in this table is copied directly from Table VI of GWTC-2 \cite{2021PhRvX..11b1053A} (O3a events) and Table IV of GWTC-3 \cite{2021arXiv211103606T} (O3b events). $^\dag$The network matched-filter SNR in this table is not to be confused with SNR$_{\rm net}$ which denotes \textit{injected} network SNR of IS1 and IS2 events.}
\label{table:offsource}
\end{table*}

As with IS1, only injections above the BW significance threshold are retained in IS2. The last column in Table \ref{table:offsource} shows the number of off-source injections that exceeds the significance threshold i.e. $\text{SNR}_\text{net}\geq\text{SNR}_\text{cut-off}=10$ for each GW event. There are four GW events with less than $25$ off-source injections (i.e. $<50\%$) satisfying the significance threshold, namely: GW190517\_055101, GW190720\_000836, GW190828\_065509 and GW200219\_094415. Since the ${\rm SNR}_{\rm net}$ of off-source injections for these four events are sampled from match-filter network SNR posteriors with medians $\lesssim 11$ (see Table \ref{table:offsource}), they are less likely to satisfy $\text{SNR}_\text{net}\geq10$. Assessments of astrophysical significance for GW events with $\leq25$ off-source injections are unreliable due to insufficient $P_{\rm FA}$ measurements. Therefore, the four events listed above are excluded from the IS2 analysis. For the remaining 18 GW events, the numbers of injections shown in the last column of Table \ref{table:offsource} include events that are more consistent with Gaussian noise than a GW signal according to BW. As discussed in Section \ref{sec:glitch_dataset}, these events are assigned $\ln\mathcal{B}_{\mathcal{S},\mathcal{G}}=-500$ to indicate low significance.

\begin{figure}[t]
\centering
\includegraphics[width=.49\textwidth]{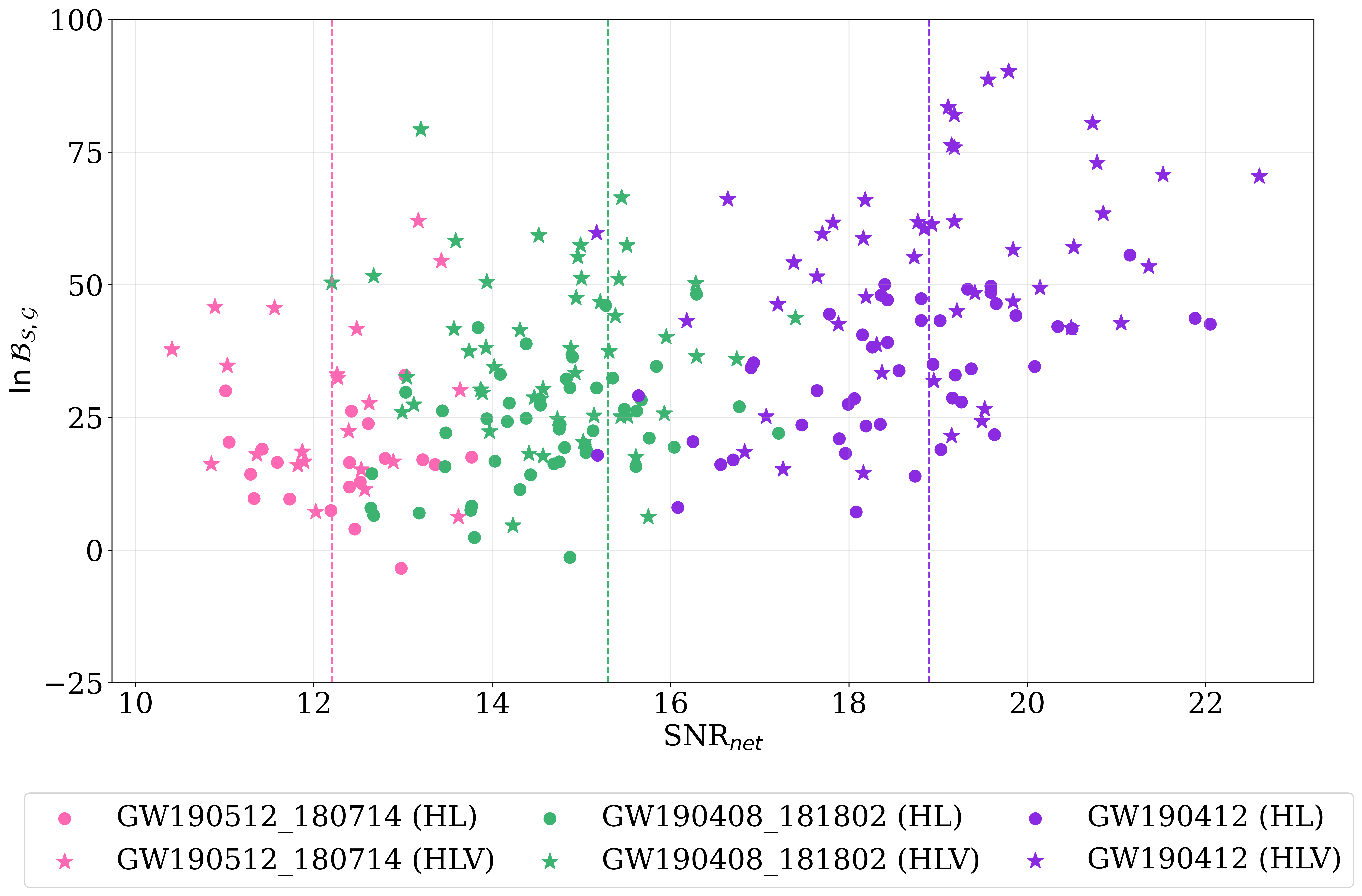}
\caption{$\ln\mathcal{B}_{\mathcal{S},\mathcal{G}}$ versus $\text{SNR}_\text{net}$ for IS2 off-source injections of GW190512\_180714 (pink), GW190408\_181802 (green) and GW190412 (purple). The vertical dashed lines in the respective colors at $\text{SNR}_\text{net}= 12.2,\,15.3,\,18.9$ indicate the median HLV network match-filter SNRs of the GW events (from Table \ref{table:offsource}). The circles and stars correspond to HL and HLV injections respectively. Gaussian-noise-like events with $\ln\mathcal{B}_{\mathcal{S},\mathcal{G}}=-500$ are not shown.}
    \label{fig:offsource_dist}
\end{figure}

Figure \ref{fig:offsource_dist} shows the distribution of off-source injections in IS2 for each GW event in different colors. To avoid clutter, we show only three arbitrarily selected events with contrasting HLV network match-filter SNRs from Table \ref{table:offsource}, namely GW190512\_180714 (pink), GW190408\_181802 (green) and GW190412 (purple). Each circle (star) data point correspond to an individual HL (HLV) injections. For each GW event in Figure \ref{fig:offsource_dist}, the off-source injection $\text{SNR}_\text{net}$ are distributed within an approximate range of $\pm 5$ from their respective median HLV network match-filter SNR, indicated by the vertical dashed lines in corresponding colors. The $\ln\mathcal{B}_{\mathcal{S},\mathcal{G}}$ also scales with $\mathcal{I}$, consistent with Ref. \cite{2021PhRvD.103f2002L}. According to Table \ref{table:offsource}, the three events in Figure \ref{fig:offsource_dist} also have comparable number of off-source injections in IS2. However, the number of injections for GW190512\_180714 (pink) is visibly lower than the other two events, because the plot excludes Gaussian-noise-like events with $\ln\mathcal{B}_{\mathcal{S},\mathcal{G}}=-500$. GW190512\_180714 has the lowest network match-filter SNR of the three events, so its offsource injections in both the HL and HLV networks also have comparably lower $\text{SNR}_\text{net}$. Hence, the BW evidences favours the Gaussian noise model more strongly than the signal model for a larger proportion ($\sim50\%$) of GW190512\_180714's offsource injections c.f. $\sim0-5\%$ for the other two events.

The comparison of BW's detection significance ($P_{\rm FA}$) between the HL and HLV networks is presented in Section \ref{sec:results_IS2}, for all 18 O3 GW events in IS2. 

\section{Background measurements}
\label{sec:backgroundnoise_BW}

In this section, we discuss the suitability of using $P_\text{FA}$ (as opposed to FAR) as a significance measure for the purpose of our analysis. We then present and discuss the noise background measurements. Using the dataset described in Section \ref{sec:glitch_dataset}, we obtain the distribution of $P_\text{FA}$ as a function of $\ln\mathcal{B}_{\mathcal{S},\mathcal{G}}$. 

$P_{\rm FA}$ is the probability that a trigger of a given detection statistic ($\ln\mathcal{B}_{\mathcal{S},\mathcal{G}}$) is a false alarm i.e. non-astrophysical.  In the context of hypothesis testing, $P_{\rm FA}$ represents the false positive rate (type I error) and is a dimensionless quantity by definition. In contrast, FAR measures the \textit{temporal} frequency of false alarms producing a detection statistic value equal to or higher than a specified GW candidate event \cite{2016PhRvD..93l2003A}. In other words, FAR is a time-average quantity which conflates BW's performance with engineering factors such as the detector glitch rate. As discussed in Section \ref{sec:glitch_dataset}, BW is not suitable for a full all-sky search of an observational dataset and is used instead to follow up triggers identified by other burst search pipelines like cWB. In this study, we measure BW's background for the HL and HLV networks using populations of background triggers arbitrarily downselected from the cWB all-sky analysis of the respective O3a time-shifted background data. Since $P_{\rm FA}$ is time-independent and marginalises over the number of triggers analysed, it relates directly to how BW is used in this study. It is therefore more appropriate to compare BW's performance between the HL and HLV networks using $P_{\rm FA}$ as a measure of detection significance\footnote{$P_{\rm FA}$ should not be confused with the definition of false alarm probability, $\rm{FAP}=1-\exp(-T_{\rm{obs}} \times \rm FAR)$ used in other analysis pipelines e.g. PyCBC \cite{Usman2016}. FAP is the probability of finding \textit{one or more} noise background events with significance equal to or higher than FAR (of a candidate event) within an observation period $T_{\rm obs}$.}. 

\begin{figure}[t]
\centering

\includegraphics[width=.49\textwidth]{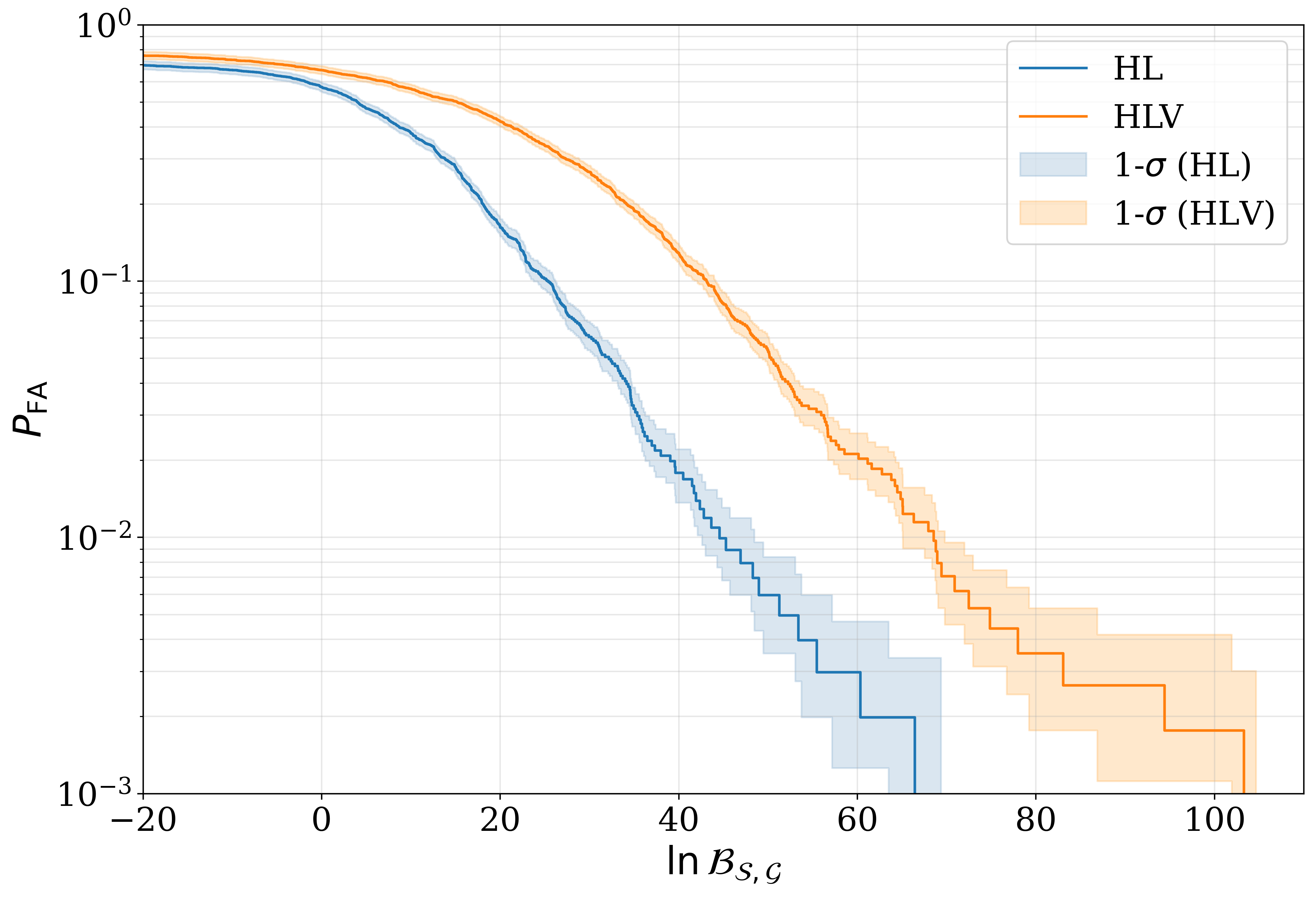}

\caption{Background measurements for the BW algorithm. The blue (orange) curve corresponds to the HL (HLV) background measured using the downselected O3a background triggers described in Section \ref{sec:glitch_dataset}. The shaded bands show the 1-$\sigma$ Poisson uncertainty regions for each network in corresponding colors.}

\label{fig:background}
\end{figure}

Figure \ref{fig:background} shows the HL and HLV network background as measured by BW with the background trigger datasets described in Section \ref{sec:glitch_dataset}.  $P_{\rm FA}$, plotted on the vertical axis, is computed as the fraction (i.e. per-trigger probability) of non-astrophysical triggers in the background exceeding the corresponding $\ln\mathcal{B}_{\mathcal{S},\mathcal{G}}$ on the horizontal axis. We restrict the plot to $\ln\mathcal{B}_{\mathcal{S},\mathcal{G}}>-20$, the range relevant to real astrophysical signals. Although not shown in Figure \ref{fig:background}, triggers with $\ln\mathcal{B}_{\mathcal{S},\mathcal{G}}<-20$ are included in the denominators for computing $P_{\rm FA}$, that is 1008 and 1134 respectively for HL and HLV. To estimate the uncertainties in our background measurements, we conventionally assume the detector noise background can be modelled as a Poisson process. The shaded regions show the 1 $\sigma$ Poisson uncertainty region for HL and HLV in corresponding colors. In Appendix \ref{app:poisson}, we show the implementation of $P_{\rm FA}$ in Poisson statistics as opposed to FAR, along with the derivation of the Poisson uncertainty regions.

The background measurements show that $P_{\rm FA}$ is higher for HLV than for HL at all $\ln\mathcal{B}_{\mathcal{S},\mathcal{G}}$ as the occurrence of background triggers increases with the number of detectors. As a result, events detected by the HLV network need to attain a higher $\ln\mathcal{B}_{\mathcal{S},\mathcal{G}}$ in order to achieve the same significance ($P_{\rm FA}$) as the HL network. For example, to achieve $P_{\rm FA}=0.1$, a HL event requires $\ln\mathcal{B}_{\mathcal{S},\mathcal{G}}=25.6$; c.f. $\ln\mathcal{B}_{\mathcal{S},\mathcal{G}}=43.1$ for HLV. Additionally, $\ln\mathcal{B}_{\mathcal{S},\mathcal{G}}$ of the HLV background triggers are higher overall compared to HL. This is because the increased trigger frequency in HLV results in the increased likelihood of coincident triggers which more closely resemble coherent signals, and are therefore recovered with higher detection statistics by BW. Furthermore, the misalignment of the Virgo detector senses a different signal polarization to the two co-aligned LIGO detectors, thus imposing a less stringent constraints on signal coherence. This reduces the efficiency of HLV in discriminating coincident glitches from signals. 

\section{BayesWave detection efficiency with BBH waveforms}
\label{sec:results_IS1}

\subsection{Constructing efficiency curves}
\label{sec:construct_effCurves}

In Ref. \cite{2016PhRvD..93b2002K}, the performance of a hierarchical pipeline consisting of cWB and BW is quantified using efficiency curves, which show the fraction of injected signal waveforms recovered above various significance thresholds. We use the same approach in this work to study the independent performance of BW. Using IS1 described in Section \ref{sec:simulated_inj}, we construct efficiency curves for the HL and HLV networks by plotting $P_{\rm det}$ as a function of $P_\text{FA}$.

As per Ref. \cite{2016PhRvD..93b2002K}, $P_{\rm det}$ is calculated as the fraction of astrophysical events recovered with detection statistic above a threshold. For BW, this threshold is set by the $\ln\mathcal{B}_{\mathcal{S},\mathcal{G}}$ corresponding to a user-selected significance i.e. $P_\text{FA}$. As noted in Figure \ref{fig:background}, the $\ln\mathcal{B}_{\mathcal{S},\mathcal{G}}$ threshold is higher for HLV than for HL at a fixed $P_\text{FA}$. The following example shows how $P_{\rm det}$ is computed for an arbitrary but representative choice $P_\text{FA}=0.2$. 

\begin{figure}[t]
\centering
\includegraphics[width=.49\textwidth]{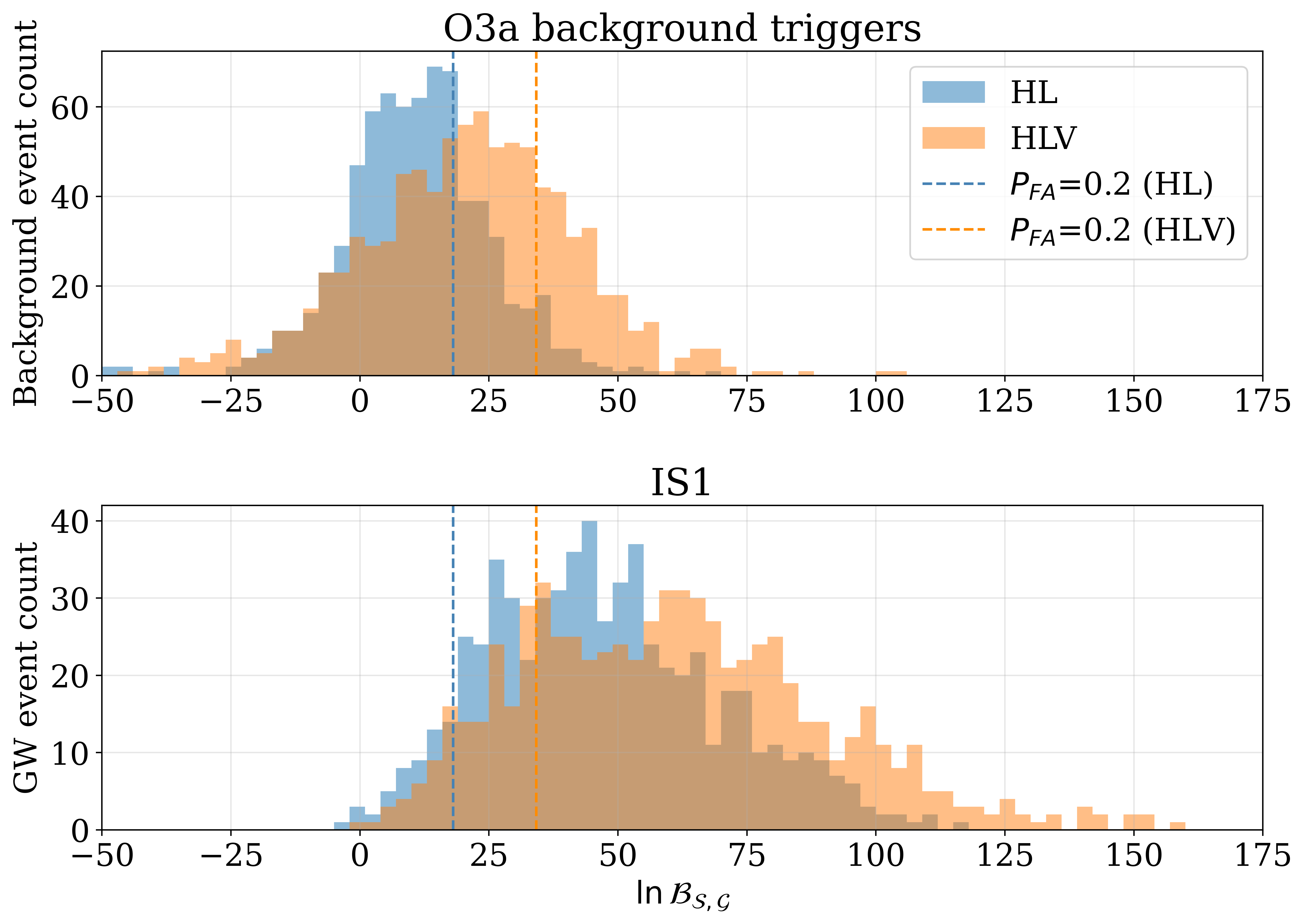}
    \caption{Worked example: computing one representative point on the efficiency curve for a significance threshold $P_{\rm FA}=0.2$. \textbf{Top panel.} Histogram of $\ln\mathcal{B}_{\mathcal{S},\mathcal{G}}$ for the O3a background triggers described in Section \ref{sec:glitch_dataset}. \textbf{Bottom panel.} Histogram of $\ln\mathcal{B}_{\mathcal{S},\mathcal{G}}$ for IS1. The HL and HLV network histograms are color-coded blue and orange respectively. In both panels, the vertical dashed lines at $\ln\mathcal{B}_{\mathcal{S},\mathcal{G}}=18.1$ (HL) and $34.2$ (HLV) indicates the threshold for $P_{\rm FA}=0.2$. The fraction of injections to the right of the thresholds in the bottom panel yields $P_{\rm det} = 0.74$ (HL) and $0.71$ (HLV).}
    \label{fig:DetEffDemo}
\end{figure}

Figure \ref{fig:DetEffDemo} shows histograms of $\ln\mathcal{B}_{\mathcal{S},\mathcal{G}}$ for the HL (blue) and HLV (orange) O3a background triggers in the top panel, and for IS1 in the bottom panel. The $\ln\mathcal{B}_{\mathcal{S},\mathcal{G}}$ thresholds for $P_\text{FA}=0.2$ is set by the background triggers in the top panel. In both panels, we indicate the thersholds by the vertical dashed lines at $18.1$ (HL, blue) and $34.2$ (HLV, orange). With the HL and HLV thresholds established, we turn to the bottom panel of Figure \ref{fig:DetEffDemo} where we compute $P_{\rm det}$ as the fraction of IS1 injections detected by HL (HLV) greater than the threshold, i.e. to the right of the blue (orange) vertical line. We find $P_{\rm det}=0.74$ and $0.71$ for HL and HLV respectively. The procedure is repeated for $P_{\rm FA}$ in the range $0 \leq P_{\rm FA} \leq 1$ to construct the efficiency curves for HL and HLV.

\subsection{Efficiency analysis}
\label{sec:effCurves}

\begin{figure}[t]
\centering
\includegraphics[width=.495\textwidth]{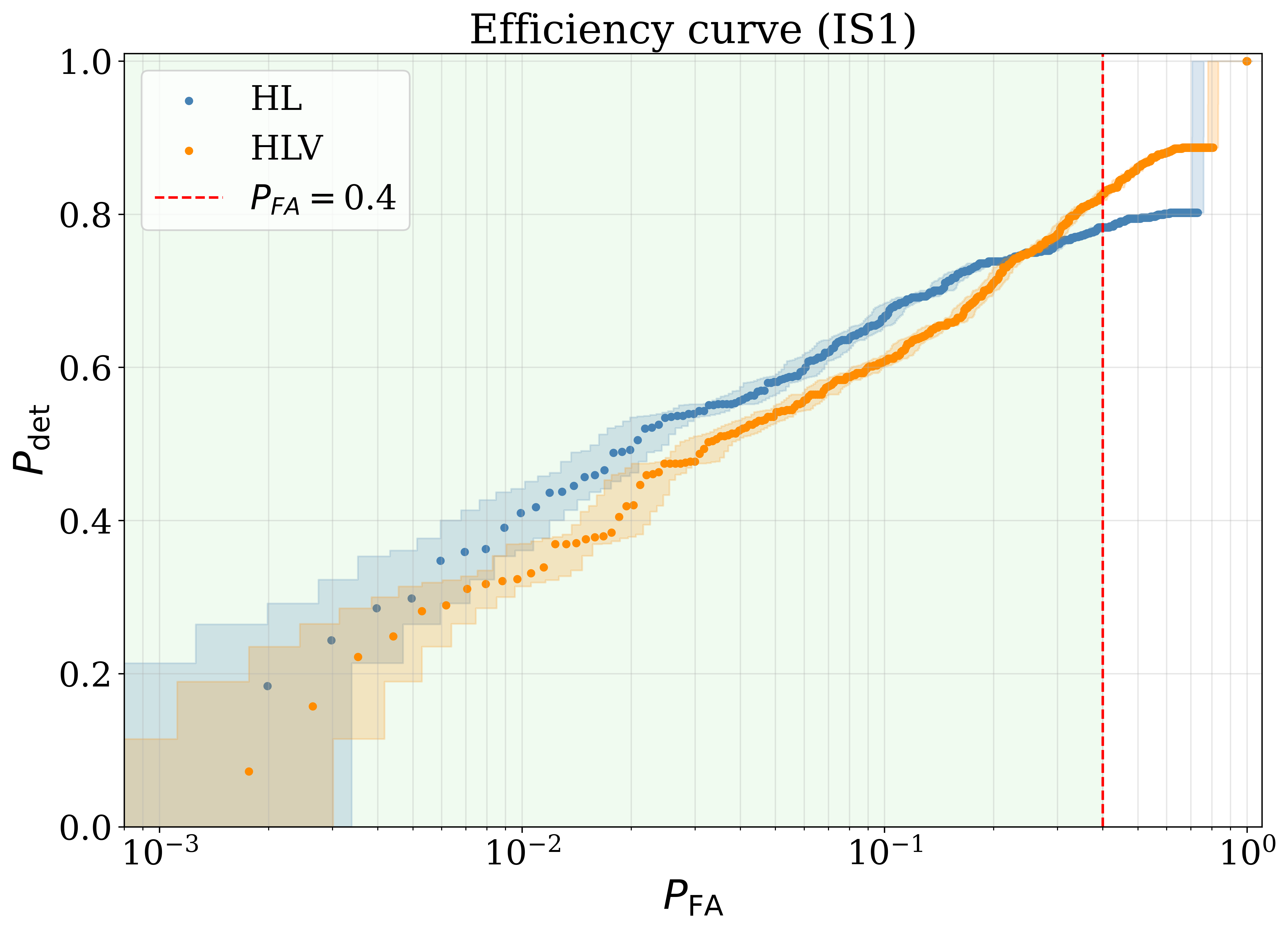}
    \caption{BW efficiency curves constructed using IS1 for the HL (blue) and HLV (orange) networks. The shaded bands with matching colors are the 1-$\sigma$ Poisson uncertainty regions for $P_{\rm FA}$, same as in Figure \ref{fig:background}. The region where $P_{\rm FA}\leq0.4$ is shaded green to indicate astrophysical relevance.}
    \label{fig:simInj_effCurve}
\end{figure}

The efficiency curves of IS1 for the characterization of BW's overall burst detection efficiency is shown in Figure \ref{fig:simInj_effCurve}. The blue and orange curves correspond to the HL and HLV networks respectively. To indicate the error margins of $P_{\rm FA}$ from the background measurements, we carry over the 1-$\sigma$ Poisson uncertainty regions onto the horizontal axis of the efficiency curves. From the background measurements, we also noted that the minimum $\ln\mathcal{B}_{\mathcal{S},\mathcal{G}}$ required to achieve a given significance reduces with increasing tolerance for $P_{\rm FA}$. Therefore the efficiency curves show that $P_{\rm det}$ increases with $P_{\rm FA}$, as more events in IS1 satisfy the reduced $\ln\mathcal{B}_{\mathcal{S},\mathcal{G}}$ threshold. The cluster of data points at $P_{\rm det}=P_{\rm FA}=1$, disjointed from the rest of the efficiency curves, is an artifact from assigning an arbitrarily low significance of $\ln\mathcal{B}_{\mathcal{S},\mathcal{G}}=-500$ to Gaussian-noise-like events. As discussed in Section \ref{sec:simulated_inj}, these events occupy $20\%$ ($11\%$) of the HL (HLV) IS1 injections. Therefore we observe a discrete jump in the fraction of recovered injections i.e. $P_{\rm det}$ of HL (HLV) from 0.80 (0.89) to 1. We also note a gap in the $P_{\rm FA}$ between the cluster of data points and the point before $P_{\rm FA}=1$. This is because the second lowest $\ln\mathcal{B}_{\mathcal{S},\mathcal{G}}$ for the HL and HLV IS1 injections are of order $-10^1$ according to Figure \ref{fig:simInj_dist}.

In order to assess the overall detection efficiency of BW with the HL and HLV networks, we focus on the region where $P_{\rm FA}$ is low enough to be astrophysically relevant. We arbitrarily define this region to be where $P_{\rm FA}\leq0.4$ as indicated by the green shading. In this region, $P_{\rm det}$ of HL is generally higher than HLV, but the opposite is true for  $P_{\rm FA} \gtrsim 0.25$. By quantifying the ratio between HL and HLV $P_{\rm det}$ for all data points in $P_{\rm FA}\leq0.4$, we find that the HL network is only 1.02 times (i.e. $2\%$) more efficient in detecting IS1 injections than the HLV network on average. Hence, there are no significant differences in BW's overall detection efficiency with a two- or three-detector configuration. 

\begin{figure}
\centering
\includegraphics[width=.495\textwidth]{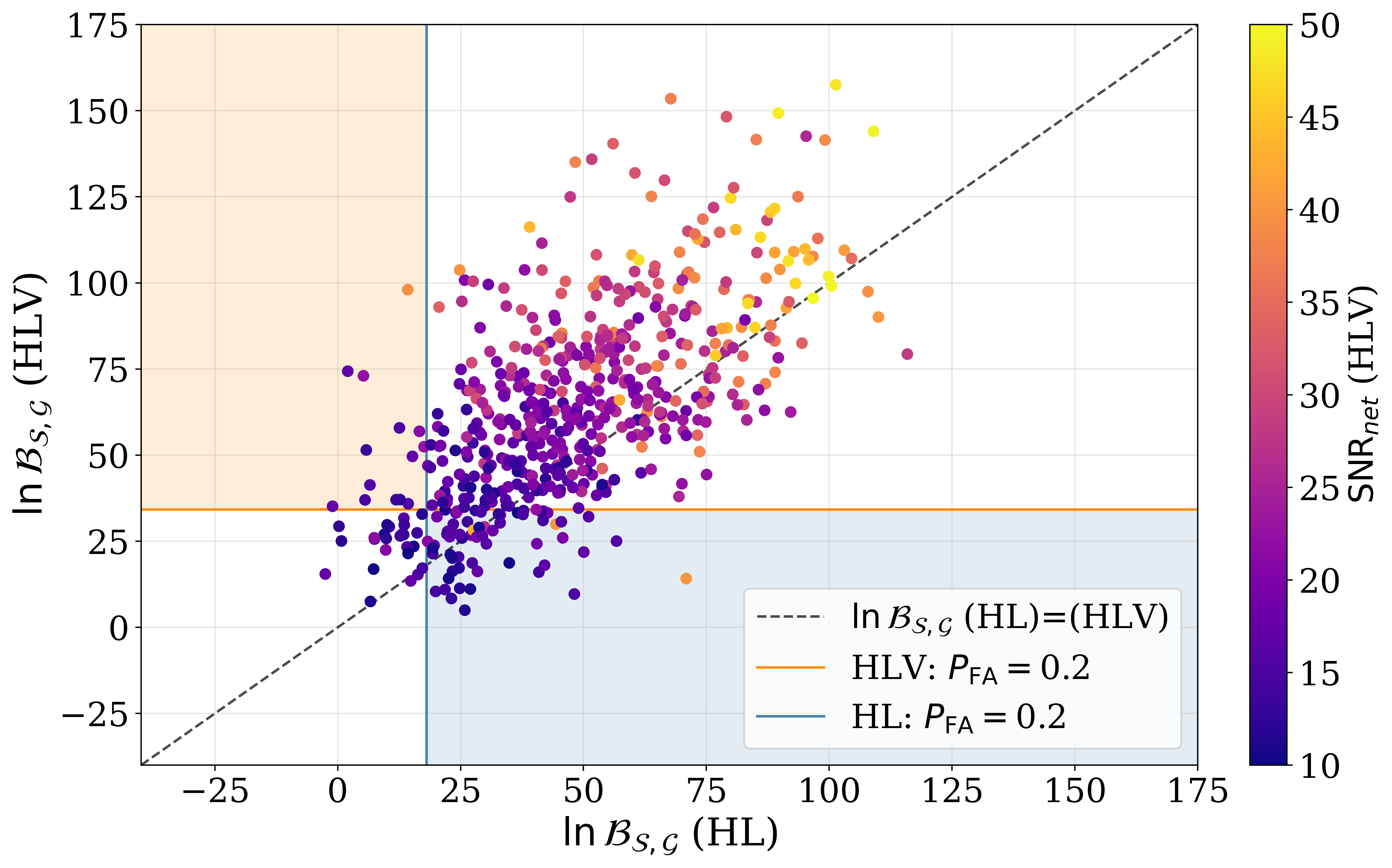}
    \caption{Log signal-to-glitch Bayes factor, $\ln \mathcal{B}_{\mathcal{S},\mathcal{G}}$ of the HLV network versus the HL network for IS1. The color bar shows SNR$_\text{net}$ in HLV for each injection. The diagonal line indicates equal $\ln\mathcal{B}_{\mathcal{S},\mathcal{G}}$ for both networks. The dashed lines at $\ln\mathcal{B}_{\mathcal{S},\mathcal{G}} \, (\rm HL)=18.1$ and  $\ln\mathcal{B}_{\mathcal{S},\mathcal{G}} \, (\rm HLV)=34.2$ indicate the thresholds for $P_\text{FA}\leq0.2$ with the respective networks. Gaussian-noise-like events with $\ln\mathcal{B}_{\mathcal{S},\mathcal{G}}=-500$ are excluded in this plot.}
    \label{fig:simInj_BSG_compare}
\end{figure}

To justify our findings, we show the event-wise comparison of $\ln\mathcal{B}_{\mathcal{S},\mathcal{G}}$ between the HL and HLV networks for IS1 in Figure \ref{fig:simInj_BSG_compare}, color-coded according to the ${\rm SNR}_{\rm net}$ of HLV\footnote{The HL and HLV network SNR$_\text{net}$ are equally representative of the ensemble SNR$_\text{net}$ of IS1 (see Figure \ref{fig:simInj_dist}). Thus we show only the HLV network SNR$_\text{net}$ in Figure \ref{fig:simInj_BSG_compare} to avoid clutter.}. The dashed diagonal line indicates where $\ln\mathcal{B}_{\mathcal{S},\mathcal{G}}$ is equal in both networks. For a specified detection significance ($P_{\rm FA}$), the plot can be divided into four quadrants by the corresponding $\ln\mathcal{B}_{\mathcal{S},\mathcal{G}}$ thresholds of the HL and HLV networks. Using $P_{\rm FA}=0.2$ again as a representative example, we indicate the HL (HLV) threshold with a blue (orange) solid line in Figure \ref{fig:simInj_BSG_compare}. The quadrants classify IS1 events based on their detectability. A successful detection in the HL (HLV) network is when the event $\ln\mathcal{B}_{\mathcal{S},\mathcal{G}}$ is higher than the detection threshold set by the blue (orange) line. By this definition, events in the top left quadrant (shaded orange) are detected by the HLV network only; the bottom right (shaded blue) by the HL network only; the top right by both networks and the bottom left by neither. We note that a fraction of events (in the blue shaded region) are only detected by HL despite having higher $\ln\mathcal{B}_{\mathcal{S},\mathcal{G}}$ in HLV. This is because a successful detection with the HLV network requires the increased $\ln\mathcal{B}_{\mathcal{S},\mathcal{G}}$ to satisfy a higher detection threshold to achieve the same significance as HL. In other words, the advantage of increased $\ln\mathcal{B}_{\mathcal{S},\mathcal{G}}$ in larger detector networks is offset by the higher detection thresholds due to the increased probability of false alarms in the background. This explains why the efficiency curves are comparable between the HL and HLV networks.

From Figure \ref{fig:simInj_BSG_compare}, we can also see that ${\rm SNR}_{\rm net}$ affects detectability. The top right quadrant contains events with overall higher ${\rm SNR}_{\rm net}$ compared to the other quadrants. That is, events with higher ${\rm SNR}_{\rm net}$ and hence higher $\ln\mathcal{B}_{\mathcal{S},\mathcal{G}}$ are more likely to be detected by both HL and HLV. The remaining IS1 events with lower ${\rm SNR}_{\rm net}$  are distributed across the other three quadrants where they fall short of at least one of the HL or HLV detectability thresholds, as indicated by the orange and blue dashed lines respectively. This is true for all $P_{\rm FA}$. We discuss the cases where events are only detected by one of the two configurations. For events detected only by HLV (orange shaded region), it is straightforward to argue that adding Virgo increases the sensitivity of the network to the signal that are too low to be detected by HL. Consequently, this increases $\ln\mathcal{B}_{\mathcal{S},\mathcal{G}}$ and boosts the detection significance past the required threshold. For the less intuitive case where events are detected only by HL (blue shaded region), we need to justify for two scenarios: (i) where $\ln\mathcal{B}_{\mathcal{S},\mathcal{G}}$ for HLV is higher than HL and (ii) vice versa. The former is discussed above. The latter suggests that the removal of Virgo boosts the signal evidence. This occurs when the addition of Virgo introduces non-Gaussian noise artifacts across the network which outweighs the sensitivity gain for the signal. These effects matter most for low ${\rm SNR}_{\rm net}$ injections. 

In summary, the efficiency curves in Figure \ref{fig:simInj_effCurve} show that BW's overall burst detection performance with the HL and HLV networks are comparable in the nominal astrophysically relevant range $P_{\rm FA}\leq0.4$. This is because the noisier detector background of the HLV offsets the advantage of increased $\ln\mathcal{B}_{\mathcal{S},\mathcal{G}}$, as revealed by the granular event analysis in Figure \ref{fig:simInj_BSG_compare}. Additionally we note that for low-$\text{SNR}_\text{net}$ injections at any given significance, adding an extra detector may tip them over or under the detection threshold unpredictably, due to a hard-to-quantify trade-off between the added noise and added sensitivity. High-$\text{SNR}_\text{net}$ injections, on the other hand, are more likely to be detected by both networks.

\section{BayesWave detection significance of O3 GW events}
\label{sec:results_IS2}

 The analysis with IS1 inferred that the overall detection efficiency of BW is comparable between the HL (two-detector) network and HLV (three-detector) network. Using IS2 described in Section \ref{sec:offsource_inj}, we conduct a consistency test for the results of IS1 by comparing BW's detection significance of O3 GW events between the two network configurations.

\begin{figure*}[t]
\centering
\includegraphics[width=0.95\textwidth]{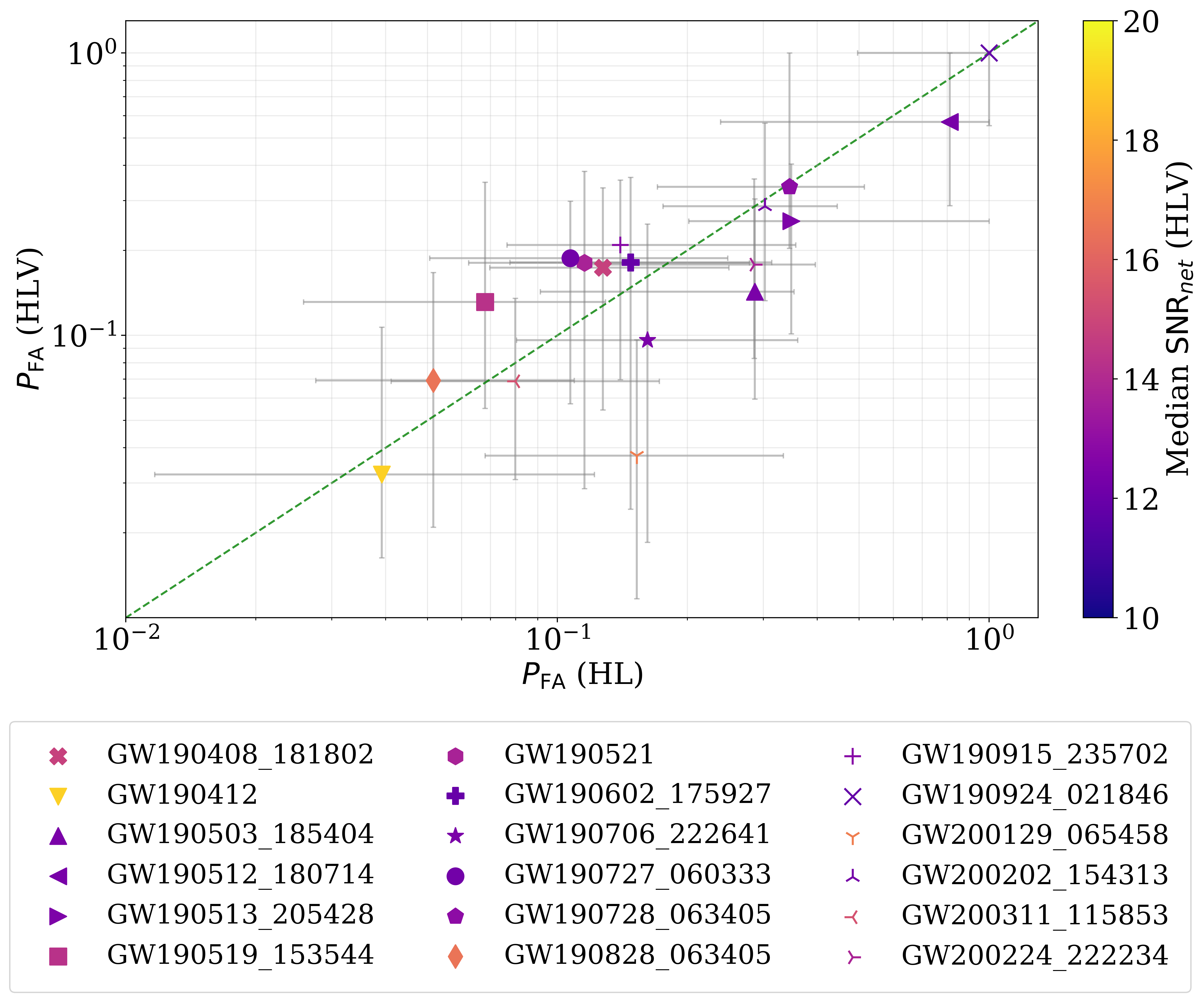}
    \caption{$P_{\rm FA}$ of the HLV network versus the HL network for O3 GW events in IS2. Each point represents a single GW event as shown in the legend and is color coded by the HLV ${\rm SNR}_{\rm net}$. The $P_{\rm FA}$ and ${\rm SNR}_{\rm net}$ shown are the medians of the off-source injections of the corresponding event; the horizontal (vertical) grey bars span the interquartile range of the HL (HLV) $P_{\rm FA}$ measurements. The diagonal line indicates equal $P_{\rm FA}$ for both networks.}
    \label{fig:O3significance}
\end{figure*}

The off-source injections in IS2 correspond to 18 independent O3 GW events. The final column of Table \ref{table:offsource} shows the number of off-source waveforms available for each event. In order to measure the detection significance of these GW events according to BW, we first quantify the significance for each off-source injection. This is done by comparing the recovered $\ln\mathcal{B}_{\mathcal{S},\mathcal{G}}$ in HL and HLV with the corresponding background measurements in Figure \ref{fig:background}. To obtain a single-valued significance measurement for each GW event, we take the median\footnote{We show the median $P_{\rm FA}$ instead of the mean, because the median value excludes any biases introduced by the Gaussian-noise like events with $P_{\rm FA}=1$, due to their arbitrarily low detection statistic $\ln\mathcal{B}_{\mathcal{S},\mathcal{G}}=-500$.} $P_{\rm FA}$ of the corresponding off-source waveforms. Figure \ref{fig:O3significance} shows the median HLV $P_{\rm FA}$ versus that of HL. We use the interquartile range (IQR), that is the range encompassing the middle 50\% of the off-source $P_{\rm FA}$ within each GW event, to represent the uncertainty in our measurements. The horizontal and vertical grey bars show the IQRs for the HL and HLV $P_{\rm FA}$ measurements respectively. We find that all data points are within close proximity of the diagonal line where $P_{\rm FA}$ is equal for HL and HLV. The size of the HL and HLV IQRs are also comparable. This suggests that BW's detection significance of O3 GW events are similar for both networks, further confirming that BW's burst detection performance with the HLV network does not exceed HL when the detector backgrounds are taken into account. 

We also note that the median $P_{\rm FA}$ increases with decreasing ${\rm SNR}_{\rm net}$, because the colors of the points darkens as one moves from the bottom left to the top right of the plot. According to Figure \ref{fig:offsource_dist}, GW events with low network match filter SNR have low ${\rm SNR}_{\rm net}$ off-source injections that are generally more consistent with Gaussian noise. Therefore the median $P_{\rm FA}$ of GW events in the top right of Figure \ref{fig:O3significance} approaches unity. This observation is consistent with Figure \ref{fig:simInj_BSG_compare}, where events with low ${\rm SNR}_{\rm net}$ and hence low $\ln\mathcal{B}_{\mathcal{S},\mathcal{G}}$ are only detectable by both HL and HLV when higher $P_{\rm FA}$ are tolerated. Furthermore, the size of the IQRs are within the same order of magnitude as the median $P_{\rm FA}$ viz. the $P_{\rm FA}$ measurement uncertainties are larger for events with lower ${\rm SNR}_{\rm net}$ off-source injections (top right corner) compared to those with higher ${\rm SNR}_{\rm net}$ (bottom left corner). The wider IQRs suggest that the increased presence of Gaussian-noise-like injections not only reduces the astrophysical significance, but also increases the uncertainty in the significance measurements for GW events with low network match filer SNR.

Altogether IS2 shows that significance measurements with BW is comparable for the HL and HLV networks, consistent with the findings of the IS1 efficiency curve analysis. We also find that $P_{\rm FA}$ and the uncertainty in its measurement increases with decreasing network match-filter SNR.


\section{Conclusion and discussion} \label{sec:conclusion}

\subsection{Summary of results}

In practice, the source-agnostic BW algorithm is used in conjunction with other search pipelines to enhance detection confidence of GW transients. In this work, however, we study the stand-alone performance of BW with expanded detector networks. Detection confidence of BW is assessed using the algorithm's detection statistics, the log signal-to-glitch Bayes factor $\ln\mathcal{B}_{\mathcal{S},\mathcal{G}}$, which measures the extent of supporting evidence for the signal model over the glitch model. A previous study shows that $\ln\mathcal{B}_{\mathcal{S},\mathcal{G}}$ increases with increasing number of detectors, $\mathcal{I}$, in a network of GW interferometers \cite{2021PhRvD.103f2002L}. However, the study did not account for the increase in glitch occurrence and the associated increase in false alarm detections, as more detectors are added to the network. This paper extends Ref. \cite{2021PhRvD.103f2002L} with the goal of determining whether BW's overall burst detection performance is enhanced or reduced as $\mathcal{I}$ increases, when the detector noise background is taken into account. This is done by measuring the noise backgrounds produced by BW and comparing the efficiency curves between the HL (two-detector) and HLV (three-detector) networks. 

We obtain the noise backgrounds measurements of BW for the HL and HLV networks by analysing non-astrophysical triggers, downselected from the cWB analysis of the O3a time-slide background data.
The background measurements show that per-trigger false alarm probability $P_\text{FA}$ is higher in the HLV network than in HL, throughout the astrophysically relevant range $\ln\mathcal{B}_{\mathcal{S},\mathcal{G}}\geq-20$. This is due to the increased likelihood of background triggers with an additional detector. We reiterate that the cWB algorithm is only used to downselect triggers for BW's background measurements, we do not investigate cWB's background and/or detection efficiency in this paper. 

For the efficiency curve analysis, we implement BW on a population of non-precessing and non-spinning phenomenological BBH waveforms (IS1) sampled from a parameter space detectable by the Advanced LIGO and Advanced Virgo detectors. IS1 is injected into segments of HL and HLV data spread out across all of O3a, to ensure comparability of the detection statistics with the background measurements. The efficiency curves plots detection efficiency, $P_{\rm det}$, of IS1 events as a function of the per-trigger false alarm probability, $P_{\rm FA}$, to characterize BW's performance over a range of significance thresholds. We find similar efficiency curves for the HL and HLV networks within a nominal significance range with plausible astrophysical implications i.e. $P_\text{FA}\leq0.4$. In other words, there are no major differences between BW's overall performance with HL and HLV. This counterintuitive finding is justified by event-wise comparison of $\ln\mathcal{B}_{\mathcal{S},\mathcal{G}}$ between the HL and HLV IS1 injections in Figure \ref{fig:simInj_BSG_compare}. The plot reveals that the advantage of increasing $\ln\mathcal{B}_{\mathcal{S},\mathcal{G}}$ with $\mathcal{I}$ is offset by the increased $P_\text{FA}$. Adding more detectors to the network increases the likelihood of noise events (i.e. false alarms). Therefore, events in larger detecter networks are required to satisfy higher detection thresholds to achieve the same significance as smaller networks. Additionally, the detectability of events by the HL and HLV networks at any given significance threshold ($P_\text{FA}$) scales with ${\rm SNR}_{\rm net}$. For events with low ${\rm SNR}_{\rm net}$, the $\ln\mathcal{B}_{\mathcal{S},\mathcal{G}}$ and hence $P_\text{FA}$ in each detector network are more sensitive to subtle changes in detector noise variation. Therefore, the addition of Virgo can unpredictably tip an event over or under the HL or HLV significance threshold.   

To check for consistency with the efficiency analysis, we separately analyse a set of O3-like CBC waveforms (IS2), otherwise referred to as off-source injections. Parameters of off-source injections are sampled from the match-filter posteriors of 18 GW events from O3. We use $P_{\rm FA}$ to quantify BW's significance for each GW event. This is evaluated by comparing the $\ln\mathcal{B}_{\mathcal{S},\mathcal{G}}$ of their respective off-source injections with the O3a background measurements. The comparison of $P_{\rm FA}$ between HL and HLV reveals that BW recovers all 18 events with similar significance from both networks. This result is consistent with the IS1 detection efficiency analysis.

Altogether, this study investigates the impact of glitches on the detection significance ($P_{\rm FA}$) and the overall performance of BW, as a function of $\mathcal{I}$. From two independent analyses with IS1 and IS2, we conclude that there are no significant differences between BW's overall burst detection performance with the HL and HLV networks. Despite the improvement in detection statistic with the addition of Virgo, the associated increase in non-astrophysical background triggers raises the detection statistic threshold which the HLV network need to attain in order to achieve the same per-trigger $P_{\rm FA}$ as HL. Therefore the HLV configuration, despite having more detectors, does not have an advantage over HL in terms of detection efficiency. Our findings are consistent with previous studies \cite{2021PhRvD.104l2004A, 2023PhRvD.107f2002S}. Although expanded detector networks improve accuracy of reconstruction and sky localisation of the GW signal, Refs. \cite{2021PhRvD.104l2004A} and \cite{2023PhRvD.107f2002S} suggest that HL rejects glitches more efficiently compared to HLV and is therefore preferred in unmodelled burst searches to maximise detection efficiency. This is because HL comprises only of the co-aligned LIGO detectors with similar sensitivities to GW polarisation components from all directions, therefore it poses more stringent constraints on signal coherence across the network. On top of that, the overall strain sensitivity of Virgo is lower than the two LIGO detectors in O3a, as shown in Figure 2 of Ref. \cite{2021PhRvX..11b1053A}. This could be another reason why the larger (HLV) network does not significantly outperform the HL network.

\subsection{Future work}
With the recently approved commissioning of LIGO-India with design sensitivity planned to match the LIGO detectors \cite{LIGOIndia}, it would be worthwhile for prospective studies on BW's detection efficiency to consider network configurations with three or more detectors of equal sensitivities.
 
Furthermore, we use trigger lists generated by the cWB algorithm from the LVK O3 all-sky burst search \cite{2021PhRvD.104l2004A} to downselect triggers for BW background measurements in this study. However, Ref. \cite{2023PhRvD.107f2002S} conducted the same search using the cWB algorithm enhanced by machine-learning (ML) which shows improved overall search sensitivity compared to the standard cWB. We therefore suggest a complementary study to follow-up on whether the BW background measurements can be improved if the triggers are downselected from the ML-enhanced cWB trigger list instead. 

While BW targets a broad range of unmodelled GW bursts, this study considers only CBC waveforms as they are the only source category detected in the LVK observing runs to date. One can generalise this study to alternative transient sources like supernovae and generic white noise bursts, but the analysis presented in this work is limited to comparing the overall trends of BW's independent performance between the HL and HLV networks. We did not study the sensitivity of BW to specific types of burst signal because BW is not used independently in practice, but rather to follow-up cWB triggers to enhance detection confidence. With promising outlooks for the ML-enhanced cWB \cite{2023PhRvD.107f2002S} and O4 in progress, future work should consider assessing the joint performance of the ML-enhanced cWB algorithm with BW for different types of burst sources as in Ref. \cite{2016PhRvD..93b2002K}.

\section*{Acknowledgements} \label{sec:acknowledgements}
    
This material is based upon work supported by NSF’s LIGO Laboratory which is a major facility fully funded by the National Science Foundation. Parts of this research were conducted by the Australian Research Council Centre of Excellence for Gravitational Wave Discovery (OzGrav), through project number CE170100004. The authors are grateful for computational resources provided by the LIGO Laboratory and supported by National Science Foundation Grants PHY-0757058 and PHY-0823459. YS. C. Lee. is supported by a Melbourne Research Scholarship and The University of Melbourne Women in Physics Award. 

We thank the members of the cWB team for producing the background triggers that we used in our analysis: Shubhanshu Tiwari, Claudia Lazzaro, Marco Drago, Francesco Salemi, Gabriele Vedovato, Sergey Klimenko. We also thank Marek Szczepa\'{n}czyk, Tyson Littenberg and Neil Cornish for their helpful comments.

\appendix
\section{BW configuration} \label{app:BWsettings}

To assist with reproducibility, we detail the BW settings for the background measurements and injection analyses. The following settings are adapted from the BW analysis used in the O3 all-sky burst search \cite{2021PhRvD.104l2004A}.

To down-select candidates from the cWB trigger list for BW's background measurements, we specify the significance threshold, $\rho_\text{threshold}=7$, as a first cut. We further reduce the dataset by keeping only a fraction of triggers satisfying $\rho>\rho_\text{threshold}$. This fraction is denoted by $X$ in the main text.

For the signal injections, we use the O3a calibrated strain data for the LIGO Hanford (H1), LIGO Livingston (L1) and Virgo (V1) detectors \cite{2023arXiv230203676T}. The frame calibration includes a noise subtraction procedure detailed in Ref. \cite{Vajente2020}. The Advanced LIGO (H1 and L1) noise subtraction targets noise from beam jitter, detector calibration lines and the main power grid line (at $60$ Hz) \cite{2019CQGra..36e5011D}. For Advanced Virgo, we use low-latency (online) strain data which includes subtraction of frequency noise from the input laser, Michelson noise from displacement of the beam splitter mirrors, amplitude noise from auxilary modulation and scattered light noise \cite{2022CQGra..39d5006A}.

For all analyses, we set the low frequency cutoff at 20 Hz by convention \cite{2020CQGra..37e5002A}. The sampling rate is set at 2048 Hz to achieve a Nyquist frequency of 1024 Hz. For PSD estimation i.e. to construct the model ${\cal N}$, we employ the \textit{BayesLine} algorithm. The BW analysis segment length is set to 4 seconds, even though GW burst signals (especially CBCs) are typically shorter. This is to ensure that detector noise is relatively stationary in analysis segment for accurate prediction of the noise spectral density with \textit{BayesLine}. Altogether, our search targets GW bursts signals with duration of milliseconds up to a few seconds, with frequencies in the $20$-$1024$ Hz frequency band of Advanced LIGO and Advanced Virgo at O3a sensitivities.

\section{Poisson noise background} \label{app:poisson}

The Poisson process models a series of randomly occurring events where the average time between events are known, but not the exact time of arrival of each event. Events modelled as Poisson process are expected to have a probability mass function given by
\begin{equation}
\centering
     P(n, \lambda) = \frac{\lambda^n \exp^{-\lambda}}{n!}.
    \label{eq:poisson}
 \end{equation}
Otherwise known as the Poisson distribution, Equation \ref{eq:poisson} measures the probability $P$ of $n$ number of events occurring within a population for a given rate parameter, $\lambda>0$. In this context, `population' refers to a group of events in a fixed temporal or spatial interval. By definition, $\lambda$ is the expected number of events in a given population, independent of the type of interval specified i.e. it is dimensionless.

\subsection{$P_{\rm FA}$ vs. $\rm FAR$ in modelling Poisson noise}
The noise background of the Advanced LIGO and Virgo detectors are modelled as a Poisson process in the standard LVK GW transient searches \cite{2019PhRvX...9c1040A, 2021PhRvX..11b1053A, 2021arXiv211103606T}. In modelling a Poisson noise background, $P_{\rm FA}$ and $\rm FAR$ play an analogous role of representing the rate of noise events, which directly influences the rate parameter $\lambda$. In the case of FAR where rate is measured in units of time, the time of observation $T_{\rm obs}$ is the interval required to obtain the expected number of noise events in the background, $\lambda = T_{\rm obs}\times \rm FAR$. Conversely, $P_{\rm FA}$ measures the noise occurrence rate in units of events. Therefore $\lambda = N_{\rm obs}\times \rm FAR$, where the interval is now given by the total number of events observed $N_{\rm obs}$. One can then show the relationship between $P_{\rm FA}$ and FAR as $P_{\rm FA}=(N_{\rm obs}/T_{\rm obs})\times \rm FAR$.

\subsection{Poisson uncertainty regions of $P_{\rm FA}$}
Since the background triggers used for BW's background measurements in Section \ref{sec:backgroundnoise_BW} are subsets of the cWB all-sky analysis of the full O3a time-shifted background of the LIGO-Virgo network, we can thereby assume the triggers obey the Poisson distribution. Consequently, we can use the standard deviation ($\sigma$) of Equation \ref{eq:poisson} to represent the error margins of our $P_{\rm FA}$ measurements. We show the derivation as follows.

In the background measurements shown in Figure \ref{fig:background}, $P_{\rm FA}$ (on the vertical axis) is computed as the fraction of background triggers recovered by BW with $\ln\mathcal{B}_{\mathcal{S},\mathcal{G}}$ exceeding the corresponding threshold, $\ln\mathcal{B}_{\mathcal{S},\mathcal{G}}^*$ (on the horizontal axis) viz.
\begin{equation}
    P_{\rm FA} = \frac{n(\ln\mathcal{B}_{\mathcal{S},\mathcal{G}}\geq\ln\mathcal{B}_{\mathcal{S},\mathcal{G}}^*)}{n_{\rm tot}},
    \label{eq:PFA}
\end{equation}
where $n_{\rm tot}$ is the total number of triggers in the background dataset (the population). The numerator is essentially the expected occurrence of events exceeding $\ln\mathcal{B}_{\mathcal{S},\mathcal{G}}^*$, hence $\lambda=n(\ln\mathcal{B}_{\mathcal{S},\mathcal{G}}\geq\ln\mathcal{B}_{\mathcal{S},\mathcal{G}}^*)$. One can then derive the 1-$\sigma$ error margin for counting the number of events $n$ exceeding $\ln\mathcal{B}_{\mathcal{S},\mathcal{G}}^*$ from the variance of the Poisson distribution:
\begin{equation}
    \sigma = \sqrt{\sum_{n=1}^\lambda (n-\lambda)^2P(n,\lambda)} = \sqrt{\lambda}.
    \label{eq:1sigma}
\end{equation}
Combining Equations \ref{eq:PFA} and \ref{eq:1sigma}, the 1 $\sigma$ Poisson uncertainty region of $P_{\rm FA}$ for a given $\ln\mathcal{B}_{\mathcal{S},\mathcal{G}}^*$ is bounded by 
\begin{equation}
    \frac{\lambda-\sqrt{\lambda}}{n_{\rm tot}}\leq P_{\rm FA} \leq \frac{\lambda+\sqrt{\lambda}}{n_{\rm tot}},
\end{equation}
as indicated by the shaded regions in Figures \ref{fig:background} and \ref{fig:simInj_effCurve}.

\begin{figure*}[t]
    \centering
    \includegraphics[width=0.49\textwidth]{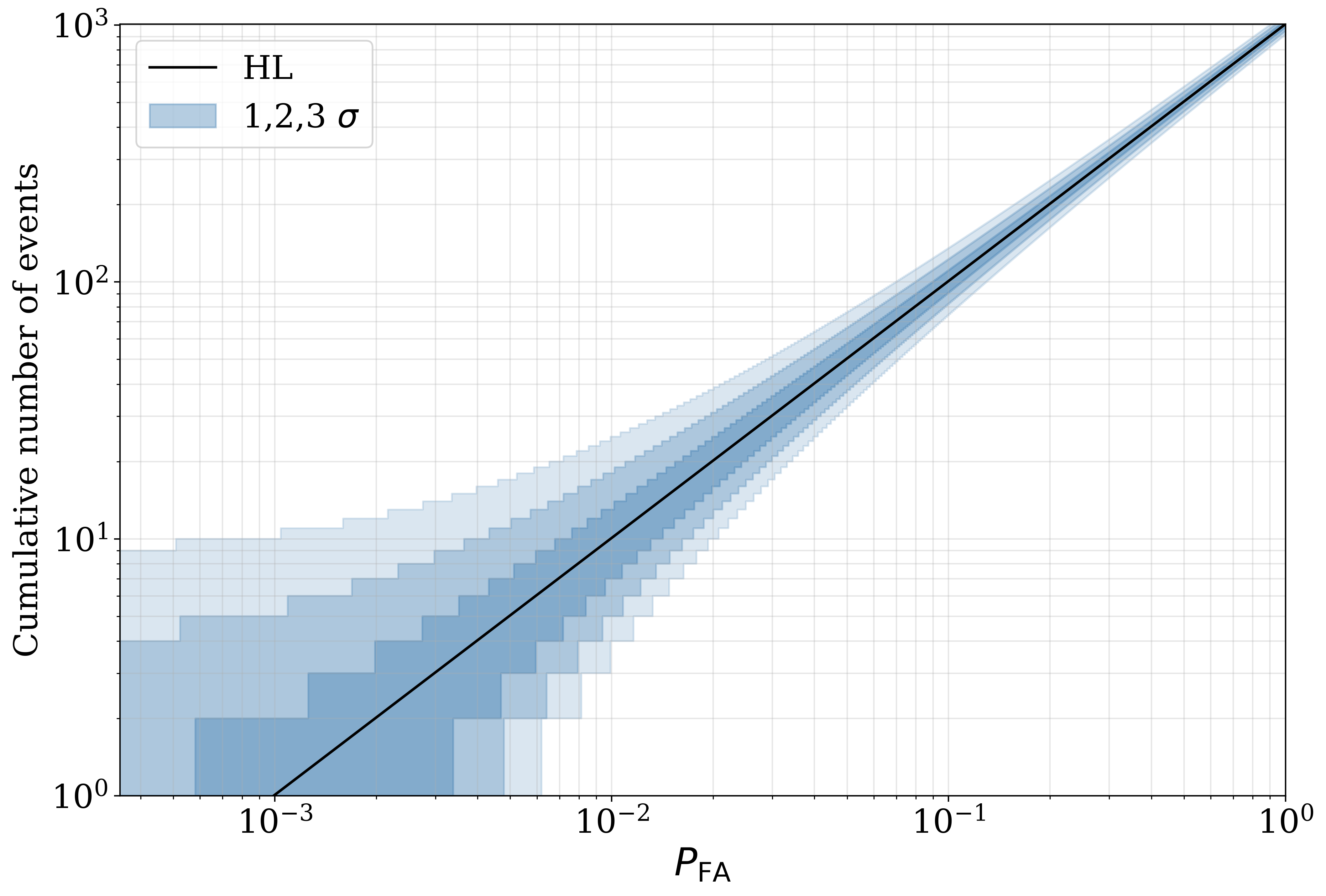}
    \includegraphics[width=0.49\textwidth]{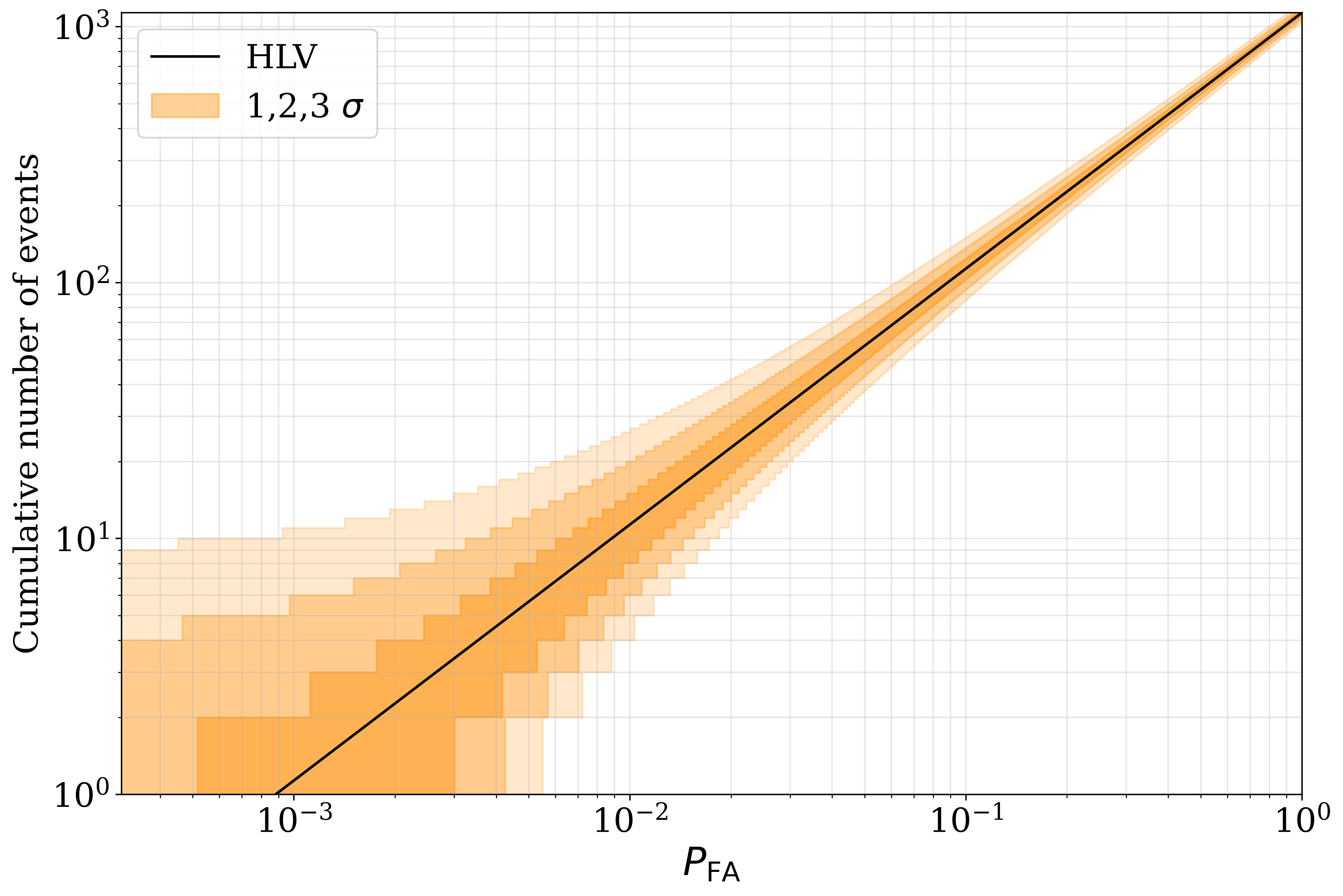}
    \caption{Cumulative number of events versus per-trigger false alarm probability, $P_{\rm FA}$ for the BW background measurements of the HL (left panel, 1008 triggers) and HLV (right panel, 1134 triggers) network.}
    \label{fig:PFA_cumulative}
\end{figure*}

To check for viability, we plot the cumulative number of triggers against $P_{\rm FA}$ in Figure \ref{fig:PFA_cumulative} and the shaded regions show the 1-, 2- and 3-$\sigma$ $P_{\rm FA}$ Poisson uncertainty regions. We compare our plots to the O3 backgrounds in Ref. \cite{2021PhRvD.104l2004A} measured with inverse FAR. Even though we use a difference quantity to measure significance, the relative extent of the shaded regions are comparable. It is therefore appropriate to use the Poisson uncertainty described above as the error margins for our $P_{\rm FA}$ measurements.

\bibliography{citations.bib}

\end{document}